\PassOptionsToPackage{dvipsnames}{xcolor}
\documentclass[sigconf,authorversion]{acmart}

\usepackage[dvipsnames]{xcolor}
\usepackage{amsmath}
\usepackage{newtxmath}
\usepackage{xspace}
\usepackage{flushend}
\usepackage{listings}
\usepackage{subcaption}
\usepackage{graphicx}
\usepackage{multirow}
\usepackage{enumitem}
\usepackage{adjustbox}
\usepackage{paralist}
\usepackage[ruled,vlined]{algorithm2e}
\usepackage[all]{nowidow}
\usepackage{stfloats}
\microtypecontext{spacing=nonfrench}

\usepackage{titlesec}
\titlespacing{\paragraph}{%
  1em}{
  0\baselineskip}{
  0.5em}
  
\DeclareMathOperator{\depth}{\mathrm{dep}}

\copyrightyear{2021}
\acmYear{2021}
\setcopyright{acmlicensed}\acmConference[SIGMOD '21]{Proceedings of the 2021 International Conference on Management of Data}{June 20--25, 2021}{Virtual Event, China}
\acmBooktitle{Proceedings of the 2021 International Conference on Management of Data (SIGMOD '21), June 20--25, 2021, Virtual Event, China}
\acmPrice{15.00}
\acmDOI{10.1145/3448016.3457263}
\acmISBN{978-1-4503-8343-1/21/06}

\begin{document}
\fancyhead{}

\newcommand{\SYS}{RisGraph\xspace}

\newcommand{\fgy}[1]{\cred{(#1 -- \textit{FGY})}}
\newcommand{\tocite}[1]{\textcolor{red}{[#1]}}
\newcommand{\toadd}{\textcolor{blue}}
\newcommand{\tofill}{\textcolor{red}{@@@}}
\newcommand{\blank}{\hspace*{0.2cm}}

\newcommand{\revision}{\textcolor{blue}}

\title[\SYS: A Real-Time Streaming System for Evolving Graphs]{\SYS: A Real-Time Streaming System for Evolving Graphs to Support Sub-millisecond Per-update Analysis at Millions Ops/s}

\author{Guanyu Feng, Zixuan Ma, Daixuan Li, Shengqi Chen, Xiaowei Zhu, Wentao Han, Wenguang Chen}
\affiliation{
    \institution{Department of Computer Science and Technology and BNRist, Tsinghua University}
    \country{}
}
\email{{fgy18, ma-zx19, li-dx17, csq20}@mails.tsinghua.edu.cn, {zhuxiaowei, hanwentao,cwg}@tsinghua.edu.cn}

\renewcommand{\shortauthors}{Guanyu Feng et al.}

\begin{abstract}

Evolving graphs in the real world are large-scale and constantly changing, as hundreds of thousands of updates may come every second. \emph{Monotonic algorithms} such as Reachability and Shortest Path are widely used in real-time analytics to gain both static and temporal insights and can be accelerated by \emph{incremental computing}. 
Existing streaming systems adopt the incremental computing model and achieve either low latency or high throughput, but not both. However, both high throughput and low latency are required in real scenarios such as financial fraud detection.




This paper presents \SYS, a real-time streaming system that provides low-latency analysis for each update with high throughput. \SYS addresses the challenge with \emph{localized data access} and \emph{inter-update parallelism}. We propose a data structure named \emph{Indexed Adjacency Lists} and use \emph{sparse arrays} and \emph{Hybrid Parallel Mode} to enable localized data access. To achieve inter-update parallelism, we propose a \emph{domain-specific concurrency control mechanism} based on the classification of \emph{safe} and \emph{unsafe} updates.

Experiments show that \SYS can ingest millions of updates per second for graphs with several hundred million vertices and billions of edges, and the P999 processing time latency is within 20 milliseconds. 
\SYS achieves orders-of-magnitude improvement on throughput when analyses are executed for each update without batching, and performs better than existing systems with batches of up to 20 million updates.

\end{abstract}



\maketitle

\section{Introduction}
\label{sec:intro}

Graphs have currently drawn broad interests in both academic and industrial communities. Real-world graphs are \emph{evolving graphs} in general, for example, social networks, financial networks, and web graphs. Evolving graphs are continuously changing~\cite{ntoulas_whats_2004, garg_evolution_2009, chen_monitoring_2009, armstrong_linkbench:_2013}, as the updates could come at a high rate, reaching tens or hundreds of thousands of updates per second~\cite{twitter_peak, qiu_real-time_2018, tmall_peak}.



Evolving graphs include both static and temporal information and can provide valuable insights with analytical algorithms. 
For example, in some e-commerce or social network analytics~\cite{erling_ldbc_2015,jiang_understanding_2016,khalil_discovering_2016,cen_trust_2020}, users are modeled into vertices, and the trust relationships between them are modeled into weighted edges. Analyzing shortest paths can discover suspicious users within short distances~\cite{khalil_discovering_2016} from known malicious users. 
Incoming interactions or transactions are converted into a series of graph updates. Updates modify the graph and change the analysis results. High throughput is essential to keep pace with incoming updates (reaching hundreds of thousands of updates per second). Meanwhile, real-time analysis is necessary to detect suspicious users and prevent fraud or harmful information from these users in time (typically within tens of milliseconds~\cite{qiu_real-time_2018}).



\textit{Monotonic algorithms}~\cite{vora_kickstarter:_2017} are commonly used~\cite{xu_fighting_2004,erling_ldbc_2015,khalil_discovering_2016,wang_querying_didi_2019} in evolving graph analytics, which include Reachability, Breadth-First Search, Shortest Path, Connected Components, and Min/Max Label Propagation. 
It requires scanning a large amount of data or even the entire graph to recompute monotonic algorithms on each snapshot of evolving graphs.
For example, the processing time of the shortest path is of the order of seconds on graphs with millions of vertices and billions of edges~\cite{kumar_graphone:_2019,shun_ligra_2013,zhu_gemini_2016}.  The idea of \emph{incremental computing} can accelerate monotonic algorithms by leveraging previous results to reduce redundant computing. 




\paragraph{\bfseries Existing Solutions.}

In recent years, several incremental graph computing models~\cite{mcsherry_differential_2013,mahmood_tornado:_2015,sengupta_graphin:_2016,vora_kickstarter:_2017} are proposed, which work well on monotonic algorithms. 
Among them, KickStarter~\cite{vora_kickstarter:_2017} and Differential Dataflow~\cite{murray_naiad:_2013,mcsherry_differential_2013} are state-of-the-art representatives. KickStarter proposes an incremental graph computing model for monotonic algorithms, while Differential Dataflow presents a generalized incremental model without graph-awareness. 
They effectively shorten the processing time of monotonic algorithms from a few seconds to milliseconds when updating a single edge on \emph{power-law graphs} with billions of edges, such as social networks~\cite{mislove_measurement_2007}, web graphs~\cite{albert_diameter_1999}, and financial graphs~\cite{qiu_real-time_2018}.




However, for graph analytics of Internet-scale, the performance of existing systems still have a large gap to fulfill the throughput requirement which would demand ingesting hundreds of thousands of updates every second. KickStarter and Differential Dataflow can handle only about one thousand updates per second if they analyze every time the graph changes (\emph{per-update analysis}). They rely on batching to trade latency for higher throughput, benefiting from larger concurrency and lower overheads. Moreover, they provide a \emph{batch-update} mode to further optimize throughput, which reduces the frequency of analyzing and produces only one aggregated final result for each batch.

Unfortunately, the high throughput of KickStarter and Differential Dataflow relies on large batches and amplifies latency, even if batch-update mode is enabled. We take Breadth-First Search (BFS) on Twitter-2010~\cite{snapnets} as an example. To meet 20 ms latency requirement (real-time analysis~\cite{qiu_real-time_2018}), the throughput of these systems is only about 1K ops/s. To provide throughput of 100K ops/s, they need to batch more than 20K updates, and the average processing time grows to more than 150 ms. Therefore, existing streaming graph systems cannot simultaneously fulfill latency and throughput requirements by batching.

Meanwhile, batch-update mode processes a batch of updates as a whole, skipping intermediate states which are potentially useful in some scenarios such as financial fraud detection~\cite{qiu_real-time_2018} and transaction conformity~\cite{barga_consistent_2007, meehan_s-store_2015}.

Compared to batching, per-update analysis is friendly to latency, produces up-to-date results, and provides the most accurate and detailed information. It only leaves one open question: how to provide high throughput in per-update analysis.

\paragraph{\bfseries Open Challenges.}
A per-update system faces challenges of combining two kinds of workloads in a fine-grained manner for each update, which include modifying the graph structure and incrementally analyzing based on the updated graph. 
Different from batch-update systems~\cite{cheng_kineograph:_2012,sengupta_graphin:_2016,vora_kickstarter:_2017,sheng_grapu:_2018,mariappan_graphbolt:_2019}, per-update analysis does not ingest multiple updates as a whole or analyzes them together. Therefore, per-update systems cannot reuse existing techniques designed for batching or benefit from batched updates, such as amortizing overheads across multiple updates. 

The goal of high throughput and low latency requires the system to efficiently conduct both kinds of workloads. When modifying the graph, it needs to apply each update to a data structure and provide an updated graph ready for analysis in a short time.
To enable real-time analysis for each update, the system requires a graph-aware design that leverages the locality of individual updates, rather than that utilizes the typical technique of entire graph scanning.
Besides, it requires a new mechanism to enable parallelism for per-update processing, which is also important to achieve high throughput without batching.

\paragraph{\bfseries Guiding Ideas.}
We propose two guiding ideas to address the challenge, \emph{localized data access} and \emph{inter-update parallelism}.

The idea of \emph{localized data access} comes from the observation that commonly used graph-aware techniques for graph streaming systems still require unnecessary entire graph scans~\cite{sengupta_graphin:_2016,vora_kickstarter:_2017,kumar_graphone:_2019,mariappan_graphbolt:_2019}. If we can avoid these scans by only accessing the necessary vertices affected by updates, we will gain much better performance, thus we propose to use data structures called \emph{Indexed Adjacency Lists} and \emph{sparse arrays} to enable localized accesses.

We further improve throughput by processing updates in parallel (\emph{inter-update parallelism}) while maintaining the per-update semantics to applications. We propose an algorithm to identify updates that can be safely executed in parallel and execute the rest of updates one by one to keep low latency, as well as atomicity, isolation, and correctness of per-update analysis.

\paragraph{\bfseries Contributions.}


We summarize our contributions as follows.
\begin{compactitem}

\item We propose a data structure for graphs named \emph{Indexed Adjacency Lists}, which provides efficient analytical performance and supports microsecond-level updates. It consists of a dynamic array of arrays and indexes of edges, to store edges in a continuous memory layout and also ensure average O(1) time complexity for each update (Section~\ref{subsec:opt_graph_storage}).

\item During incremental computing, we track active vertices and updated results with \emph{sparse arrays} to eliminate redundant overheads of scanning all the vertices and achieve localized data access. For better analytical performance, we further propose \emph{Hybrid Parallel Mode}, which adaptively uses \emph{edge-parallel} and \emph{vertex-parallel} strategies through a linear classifier (Section~\ref{subsec:opt_computing_engine}).


\item We propose a \emph{domain-specific concurrency control mechanism} to support parallelism among multiple updates with low overheads, thus improve throughput. It leverages the incremental model and intermediate data structures to identify non-conflicting updates before processing them (Section~\ref{sec:opt_throughput}).
\end{compactitem}

Based on the above core ideas, we design and implement a real-time graph streaming system for monotonic algorithms called \SYS.
For graphs with hundreds of millions of vertices and billions of edges, \SYS can ingest millions of updates per second on a single commodity machine for monotonic algorithms like Breadth-First Search (BFS), Single Source Shortest Path (SSSP), Single Source Widest Path (SSWP) and Weakly Connected Component (WCC). Meanwhile, \SYS ensures that more than 99.9\% updates can be processed within 20 milliseconds without breaking per-update analysis semantics. It provides 2-4 orders of magnitude improvement over existing solutions for per-update analysis. Besides, it performs better than KickStarter and Differential Dataflow in batch-update scenarios with up to 20M updates per batch.

\section{High-Level Architecture}
\label{sec:architecture}


\begin{figure}[b]
\centering
\includegraphics[width=0.85\columnwidth]{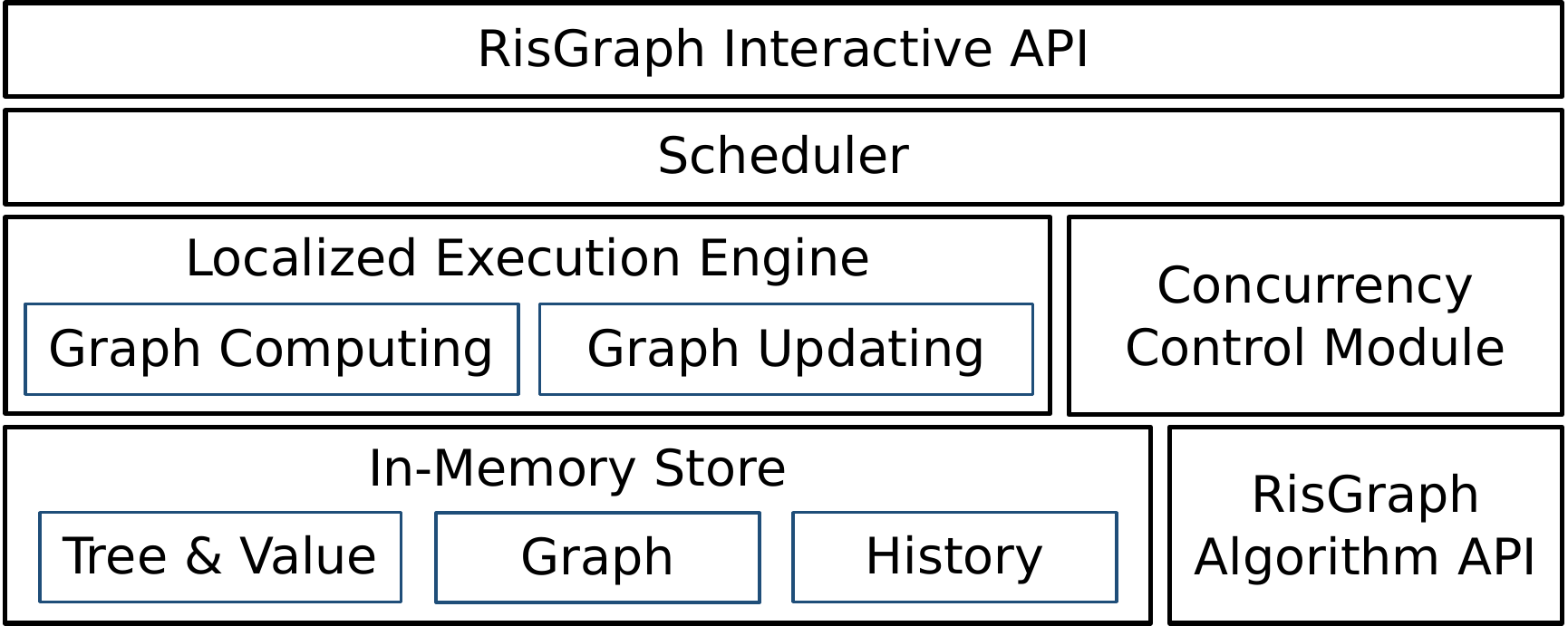}
\caption{High-level architecture of \SYS}
\label{fig:architecture}
\vspace{-1em}
\end{figure}

\SYS adopts a four-tier architecture, as shown in Figure~\ref{fig:architecture}.

The top layer is an interactive interface that allows users to interact with \SYS in a fine-grained manner. 
Users submit graph updates to \SYS and receive analysis results for each update.

A \emph{scheduler} stands below the interface to control the execution of updates from multiple clients. It monitors processing latency and dynamically schedules updates based on statistics. The goal of the scheduler is to fulfill predefined expected tail latency and achieve balanced trade-off between throughput and latency.

At the core of \SYS is the \emph{localized execution engine}, which processes each update from the scheduler through \emph{localized data accesses} (see Section~\ref{sec:opt_latency}). It includes a \emph{graph updating engine} and a \emph{graph computing engine}. 
Graph updating engine can apply updates to the graph concurrently. After updating the graph, the graph computing engine performs parallel incremental computing to synchronize analysis results based on the latest graph structure.

Independent from the execution engine, a standalone \emph{concurrency control module} ensures per-update analysis semantics, as well as correctness, atomicity and isolation. To achieve \emph{inter-update parallelism}, we propose a domain-specific mechanism which picks out non-conflicting updates before execution (see Section~\ref{sec:opt_throughput}).

An \emph{in-memory store} as an underlying layer manages all necessary data, including \emph{graph store}, \emph{tree and value store}, and \emph{history store}. 
Graph store maintains the current graph and supports efficient modification and analysis. To support incremental computing, tree and value store handles the latest and temporary computing states for each vertex. History store traces all result changes by versioning, to generate consistent result views for each update.

Finally, monotonic algorithms define the analysis tasks maintained by \SYS. \emph{Algorithm API} in \SYS makes graph monotonic algorithms easy to write.
The execution engine and concurrency control module leverage Algorithm API to perform incremental computing.

\begin{table}[t]
\caption{\SYS's Algorithm API and Interactive API}
\vspace{-1em}
\label{tab:UDF_API}
\center
\ttfamily
\resizebox{\columnwidth}{!}{
\setlength{\tabcolsep}{0.2em}
\begin{tabular}{rll}
\hline
\textcolor{Blue}{\textbf{\lstinline|init_val|}} &\lstinline|(vid)| & $\rightarrow$ \lstinline|init_value| \\ 
\textcolor{Blue}{\textbf{\lstinline|gen_next|}} &\lstinline|(edge, src_value)|& $\rightarrow$ \lstinline|next_value| \\ 
\textcolor{Blue}{\textbf{\lstinline|need_upd|}} &\lstinline|(vid, cur_value, next_value)|& $\rightarrow$ \lstinline|is_needed| \\ \hline
\\\hline
\textcolor{Blue}{\textbf{\lstinline|ins/del_edge|}} &\lstinline|(edge)| & $\rightarrow$ \lstinline|version_id| \\
\textcolor{Blue}{\textbf{\lstinline|ins/del_vertex|}} &\lstinline|(vertex_id)| & $\rightarrow$ \lstinline|version_id| \\
\textcolor{Blue}{\textbf{\lstinline|txn_updates|}} &\lstinline|(updates)|& $\rightarrow$ \lstinline|version_id| \\
\textcolor{Blue}{\textbf{\lstinline|get_value|}} &\lstinline|(version_id, vertex_id)|& $\rightarrow$ \lstinline|value|  \\
\textcolor{Blue}{\textbf{\lstinline|get_parent|}} &\lstinline|(version_id, vertex_id)|& $\rightarrow$ \lstinline|edge|  \\
\textcolor{Blue}{\textbf{\lstinline|get_current_version|}} &\lstinline|( )|& $\rightarrow$ \lstinline|version_id| \\
\textcolor{Blue}{\textbf{\lstinline|get_modified_vertices|}} &\lstinline|(version_id)|& $\rightarrow$ \lstinline|vertex_ids|  \\
\textcolor{Blue}{\textbf{\lstinline|release_history|}} &\lstinline|(version_id)|&  \\
\hline
\end{tabular}
}

\end{table}

\begin{table}[t]
\caption{Implementation of algorithms with Algorithm API}
\vspace{-1em}
\label{tab:UDF_use_case}
\center
\resizebox{\columnwidth}{!}{
\setlength{\tabcolsep}{0.15em}
\begin{tabular}{l|c|c|c|c}
\hline          &         \multicolumn{1}{c|}{BFS}          & \multicolumn{1}{c|}{SSSP}         & \multicolumn{1}{c|}{SSWP}         & \multicolumn{1}{c}{WCC} \\ \hline
\textcolor{Blue}{\textbf{\texttt{\lstinline|init_val|}}} &
    $    
        \left\{\begin{matrix}
            0 & \texttt{vid} = \textit{root}\\ 
            \infty & \texttt{vid} \neq \textit{root}
        \end{matrix}\right.
    $&   
    $    
        \left\{\begin{matrix}
            0 & \texttt{vid} = \textit{root}\\ 
            \infty & \texttt{vid} \neq \textit{root}
        \end{matrix}\right.
    $&
    $    
        \left\{\begin{matrix}
            \infty & \texttt{vid} = \textit{root}\\ 
            0 & \texttt{vid} \neq \textit{root}
        \end{matrix}\right.
    $&    
    \texttt{vid}
    \\ 

\textcolor{Blue}{\textbf{\texttt{\lstinline|gen_next|}}}                 &
$    \texttt{src\_val}+1           $        & 
$  \texttt{src\_val} + \texttt{e.data}       $          &
$  \min \left \{ \texttt{e.data}, \texttt{src\_val} \right \}    $                &
$  \texttt{src\_val}          $

\\  

\textcolor{Blue}{\textbf{\texttt{\lstinline|need_upd|}}} & 
$\texttt{next} < \texttt{cur} $               &
$\texttt{next} < \texttt{cur} $                 &
$\texttt{next} > \texttt{cur} $                 &
$\texttt{next} < \texttt{cur} $

\\ \hline
\end{tabular}
}
\vspace{-1em}
\end{table}

\paragraph{\bfseries Algorithm API.}

\SYS focuses on monotonic algorithms. Monotonic algorithms employ intermediate values to monotonically approximate and finally reach accurate results from initial values. 
After applying edge insertions, computing can start from current results instead of initial values. These results are valid intermediate values and save computation by giving a closer approximation.
KickStarter~\cite{vora_kickstarter:_2017} further discovers that all known monotonic graph algorithms, such as Reachability, Shortest Path, Weakly Connected Components, Widest Path, and Min/Max Label Propagation, can be incrementally computed by maintaining a \emph{dependency tree} (or forest). The dependency tree shows a vertex's result depends on its parent and the edge between them. Deleting an edge on the dependency tree will invalidate the subtree rooted at the destination. \emph{Trimmed approximation} technique proposed by KickStarter can generate valid approximations for invalidated vertices.



\SYS adopts the dependency tree model and the trimmed approximation technique.
Similar to KickStarter, \SYS provides user-friendly API to describe monotonic algorithms as shown in the upper part of  Table~\ref{tab:UDF_API}. \textcolor{Blue}{\textbf{\texttt{\lstinline|init_val|}}} defines initial values for each vertex. \textcolor{Blue}{\textbf{\texttt{\lstinline|gen_next|}}} uses an edge and its source vertex to generate a next possible value for its destination vertex. \textcolor{Blue}{\textbf{\texttt{\lstinline|need_upd|}}} decides whether a vertex's value should be updated. Table~\ref{tab:UDF_use_case} shows implementation of Breadth-First Search, Single Source Shortest Path, Single Source Widest Path and Weakly Connected Component.




\paragraph{\bfseries Interactive API.}

\SYS supports vertex/edge insertions and deletions, with guarantees of correctness, atomicity, isolation and per-update analysis semantics. For each operation, \SYS responds with a view of the results afterwards. Result views are versioned snapshots maintained by \SYS, to enable consistent read operations on different vertices. Optionally, \SYS provides \emph{durability} with write-ahead logs (WAL) and also \emph{sequential consistency} for better user-friendliness. Besides single edge/vertex updates, \SYS can also handle transactions (see Section ~\ref{sec:opt_throughput}).

The lower part of Table~\ref{tab:UDF_API} lists \SYS's Interactive API. Users call \textcolor{Blue}{\textbf{\texttt{\lstinline|ins / del_edge|}}} and
\textcolor{Blue}{\textbf{\texttt{\lstinline|ins / del_vertex|}}} to send updates to \SYS. 
\textcolor{Blue}{\textbf{\texttt{\lstinline|txn_updates|}}} is the API to describe a transaction or an atomic batch of updates, which is a contiguous sub-sequence of a stream that must be treated as an indivisible unit~\cite{meehan_s-store_2015}.
After \SYS processes an update or a transaction, it will return a version ID of the result snapshot. Users can get consistent results and dependency trees according to the version ID and vertex ID by \textcolor{Blue}{\textbf{\texttt{\lstinline|get_value|}}} and \textcolor{Blue}{\textbf{\texttt{\lstinline|get_parent|}}}. \SYS supports querying the current version (\textcolor{Blue}{\textbf{\texttt{\lstinline|get_current_version|}}}) and vertices that have been modified in any specific version (\textcolor{Blue}{\textbf{\texttt{\lstinline|get_modified_vertices|}}}). \SYS holds historical snapshots for each session, and requires users to actively report the latest unused versions for garbage collections (GC). \textcolor{Blue}{\textbf{\texttt{\lstinline|release_history|}}} marks previous snapshots before its version ID as no longer used for its session.

\begin{figure}[t!]
\centering
\includegraphics[width=0.99\linewidth]{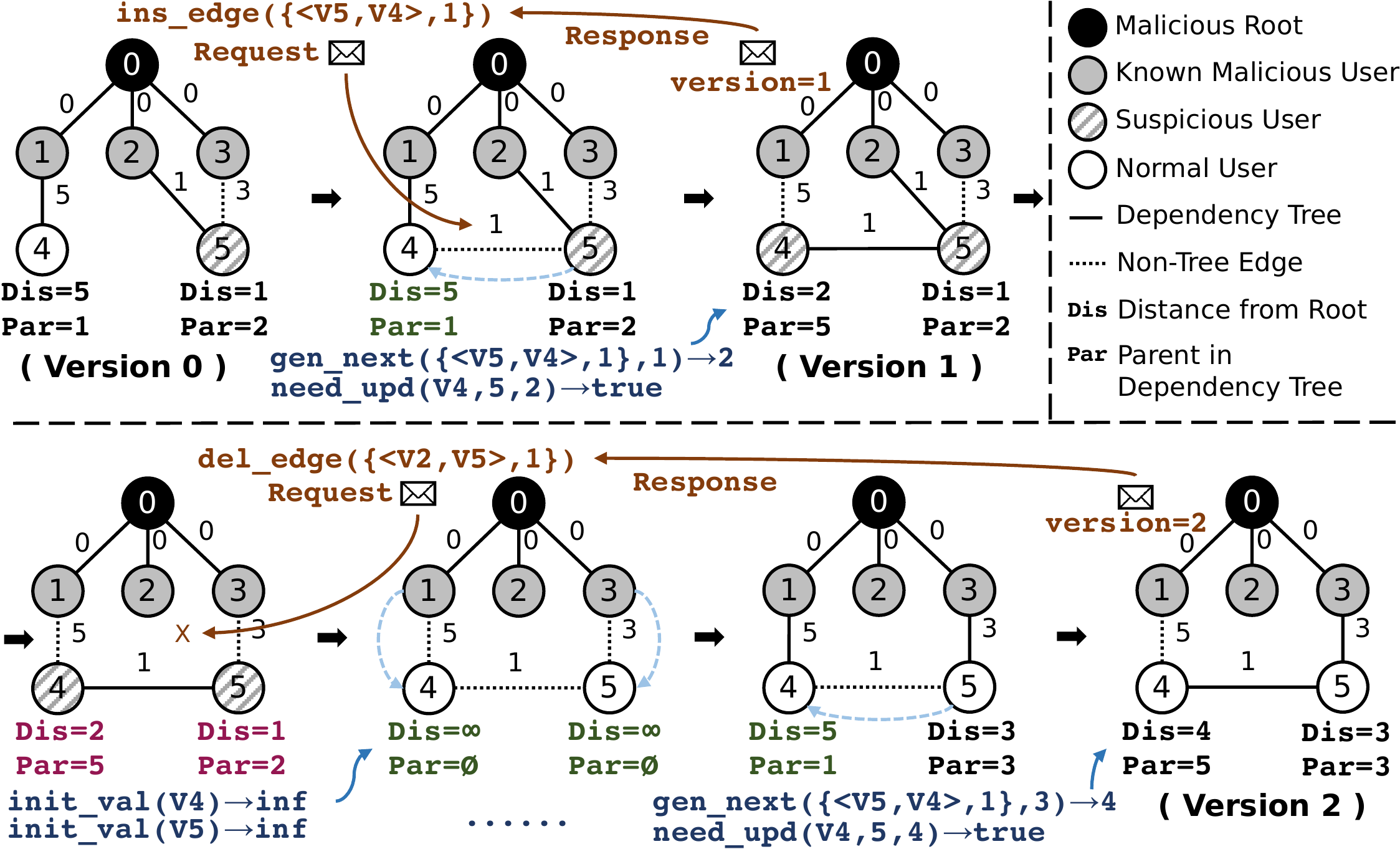}
\caption{Detecting suspicious users by SSSP, whose distance from known malicious users are defined to be within 2}
\label{fig:overall_example}
\vspace{-1em}
\end{figure}

\paragraph{\bfseries Example of \SYS API.}
Figure~\ref{fig:overall_example} shows an example of detecting suspicious users by Shortest Path algorithm. Two interactive API calls generate three result versions. At the bottom, Algorithm API calls incrementally maintain the shortest path to Vertices~4 and 5. The example also shows that detailed information provided by per-update analysis is important. It is able to find Vertex~4 suspicious when an edge is inserted in Version~1 under per-update analysis. However, it would miss the detection with batch-update mode if the system skips Version~1 and only checks Version~2.

\section{Localized Data Access}
\label{sec:opt_latency}
Different from batch-update systems, per-update analysis systems need to perform analysis for each update, thus cannot amortize the overhead of batch processing among multiple updates.
To address this challenge, we design \SYS with \emph{localized data access}, which means only accessing necessary vertices, including the vertices with results to update and the neighbours of the updated vertices when calculating new results. The rationale behind localized data access is that incremental computing only needs to access part of the vertices, but improper data structures would require partial computing to access the entire graph. The system should use proper data structures to avoid unnecessary operations.

\SYS provides 2-4 orders of magnitude improvement in our evaluation by both utilizing graph awareness and localizing data access. We make new design for the graph store and the graph computing engine, to eliminate redundant scanning. 



\subsection{Graph Store}
\label{subsec:opt_graph_storage}
For per-update analysis, graph store needs to handle each individual update and provide an updated graph ready for efficient analysis in a short time. 
Existing literature~\cite{zhu2019livegraph,kumar_graphone:_2019}
has shown that using an array of arrays to store adjacency lists can support updates and provide comparable computing performance of compressed sparse row (CSR). Graph streaming systems such as GraphOne~\cite{kumar_graphone:_2019}, GraphBolt~\cite{mariappan_graphbolt:_2019} and KickStarter also adopt an array of arrays as their graph store. However, their data structures cannot satisfy localized data access because they scan all the vertices when applying updates, even if processing a single update. With bloom filters, LiveGraph~\cite{zhu2019livegraph} supports fine-grained edge insertions well but suffers from scanning edges on \textit{hubs} (the high-degree vertices) when deleting edges.

\begin{figure}[t]
\centering
\includegraphics[width=\columnwidth]
{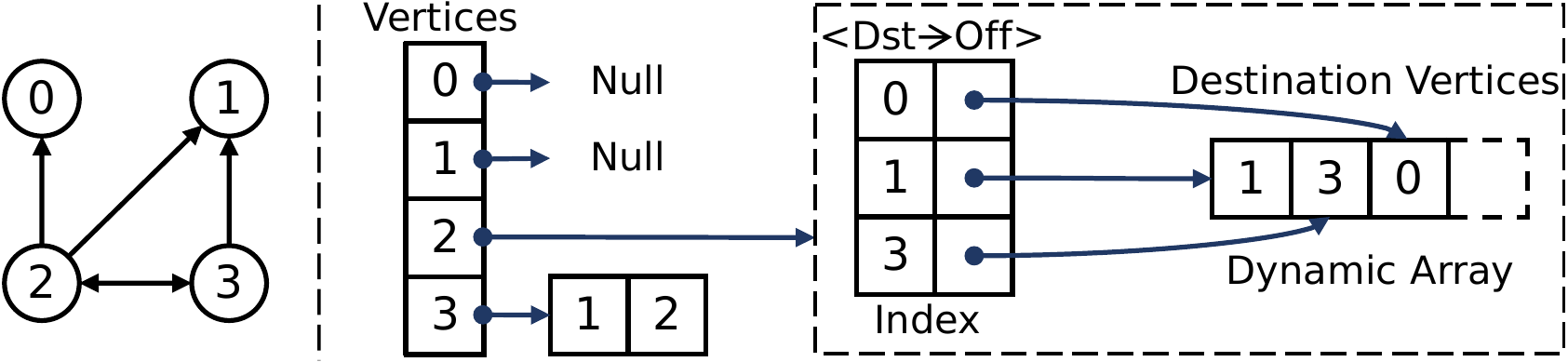}
\caption{An example of \SYS's \emph{Indexed Adjacency Lists}}
\label{fig:data_structure_example}
\vspace{-0.5em}
\end{figure}

\paragraph{\bfseries Indexed Adjacency Lists.}
\SYS proposes a data structure named \emph{Indexed Adjacency Lists} as shown in Figure~\ref{fig:data_structure_example}, which uses an array of arrays to store edges for efficient analyzing and also maintains indexes of edges to address the shortcomings of the existing approaches. 
In \SYS, each vertex has a dynamic array (doubling capacity when full) to store its outgoing edges, including destination vertex IDs and edge data.
Arrays ensure that all the outgoing edges of a vertex are continuously stored, which is critical for efficient analysis~\cite{zhu2019livegraph}.
However, arrays suffer from edge lookups by scanning for fine-grained updates. To accelerate lookups, \SYS maintains Key-Value pairs of \texttt{$\langle$DstVid$\rightarrow$Offset$\rangle$} for edges, indicating edge locations in arrays. Indexes are created only for vertices whose degree is larger than a threshold, providing a trade-off between memory consumption and lookup performance by filtering out low-degree vertices in power-law graphs (see Section~\ref{subsubsec:graph_storage}).




\SYS uses Hash Table as the default indexes because our data structure with Hash Table provides an average $\text{O}(1)$ time complexity for each update. Also, indexes do not hurt \SYS's analyzing performance because the graph computing engine can directly access adjacency lists without involving indexes. 

\begin{figure}[t]
\centering
\begin{minipage}{0.48\columnwidth}
	\centering
    \includegraphics[width=\columnwidth]{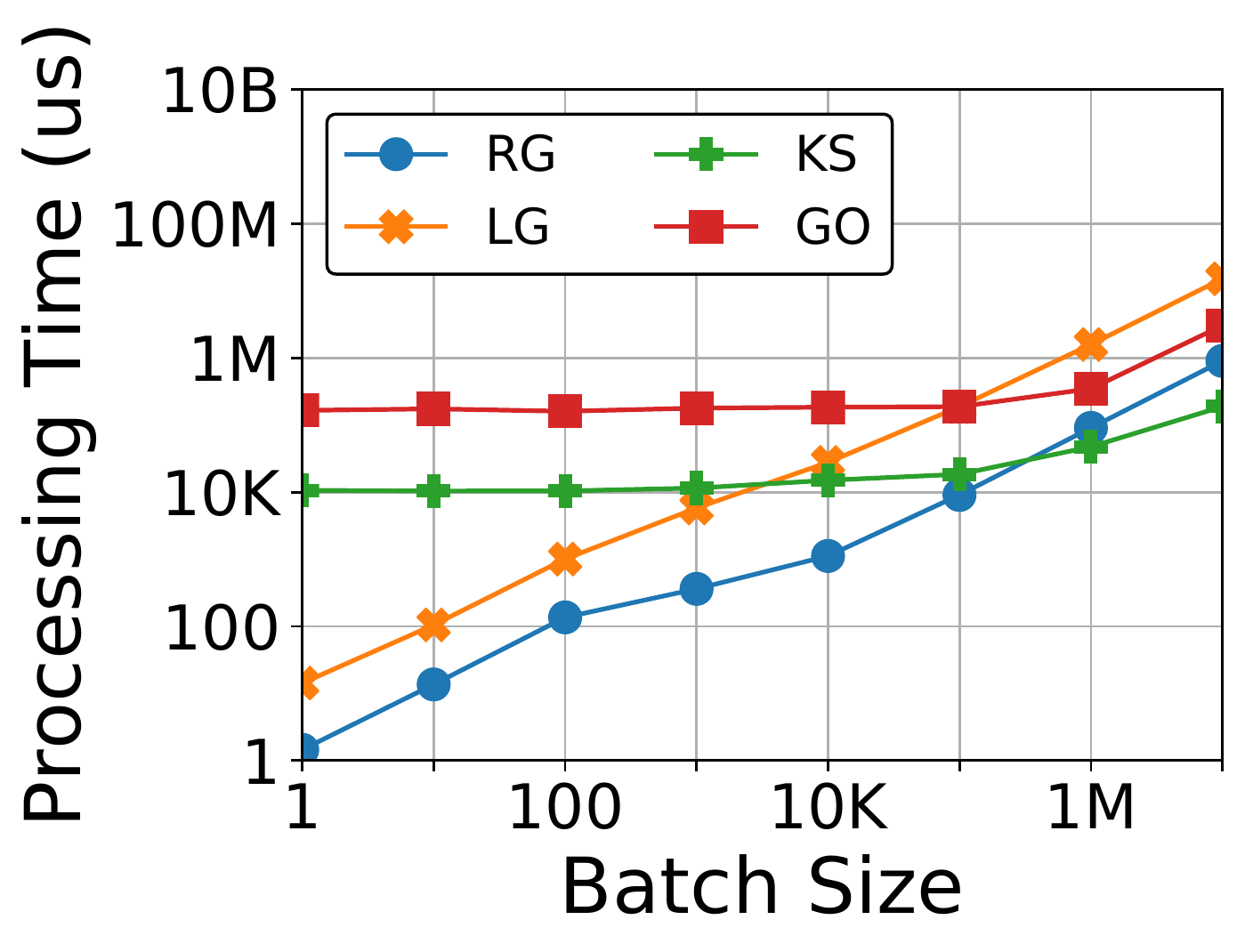}
    \subcaption{Edge insertions
    \label{fig:perf_twitter_additions}}
\end{minipage}
\hspace{0.02\columnwidth}
\begin{minipage}{0.48\columnwidth}
	\centering
    \includegraphics[width=\columnwidth]{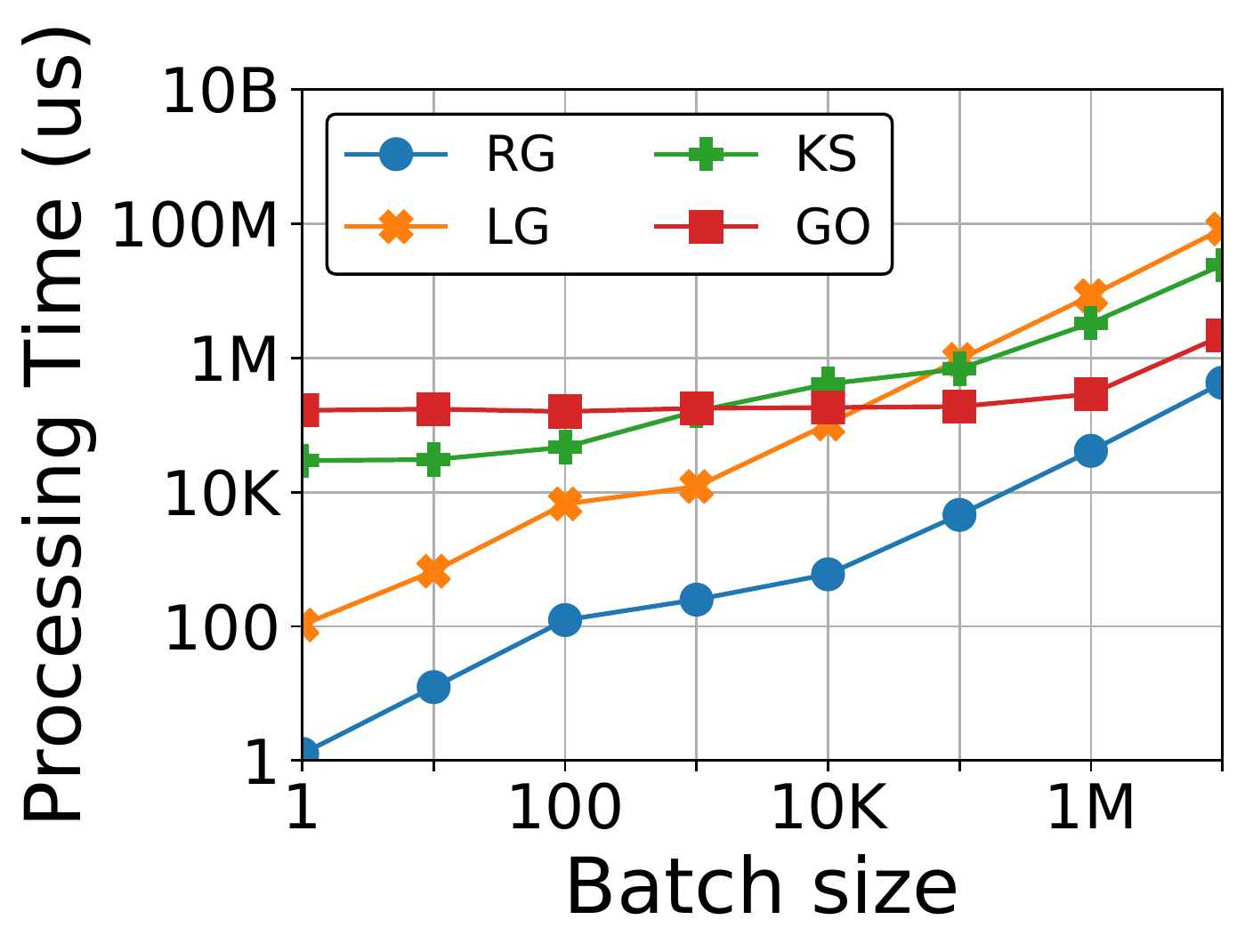}
    \subcaption{Edge deletions \label{fig:perf_twitter_deletions}}
\end{minipage}
\vspace{-0.5em}
\caption{The ingesting time of \SYS(RG), KickStarter(KS), LiveGraph(LG), and GraphOne(GO)
\label{fig:perf_twitter_datastructure}}
\vspace{-1em}
\end{figure}

Figure~\ref{fig:perf_twitter_datastructure} shows the performance of graph store when processing on Twitter-2010. The experimental setup is the same as Section~\ref{sec:evaluation}. For insertions and deletions of an edge, the average latency for updating the graph store of \SYS is several microseconds. Because of the significant improvement in performance, it is worthy of adding indexes in \SYS's data structure, at the cost of about 3.25x memory footprints of the raw data.

Under single edge updates, \SYS's graph store outperforms KickStarter and GraphOne more than thousands of times, thanks to the elimination of overheads on scanning vertices.
Compared to LiveGraph, \SYS reduces the average latency of edge insertions by 89.7\% and deletions by 98.8\%.
The indexes of \SYS can directly locate edges without scanning, which is friendly to deletions and also solves false-positive issue in LiveGraph (e.g. scanning average 541 edges per edge insertion on Twitter-2010). 
For batch updates, it is interesting to see that \SYS performs better than other systems when the batch size does not exceed 100K.


\subsection{Graph Computing Engine}
\label{subsec:opt_computing_engine}
After updating the graph, the incremental computing can be expressed by the vertex-centric model~\cite{malewicz_pregel_2010}. Some optimizations on vertex-centric frameworks have been widely discussed, for example, various methods in parallelization schemes, traversal directions, and data layouts. They target the optimization of computing on the whole graph, so the designs and trade-offs do not fully meet incremental 
graph computing. We focus on localizing data access to fit in the incremental computing scenario.

\paragraph{\bfseries Sparse Arrays.}
In vertex-centric graph computing, the most commonly used operation is the push operation~\cite{beamer_direction-optimizing_2012,besta_push_2017}. 
It iteratively performs on active vertices, whose states need to propagate. In each iteration, the push operation traverses the edges of the active vertices, updates the state of the destination vertices, and activates some destination vertices for the next iteration.

\begin{figure}[t]
\centering
\includegraphics[width=0.83\columnwidth]
{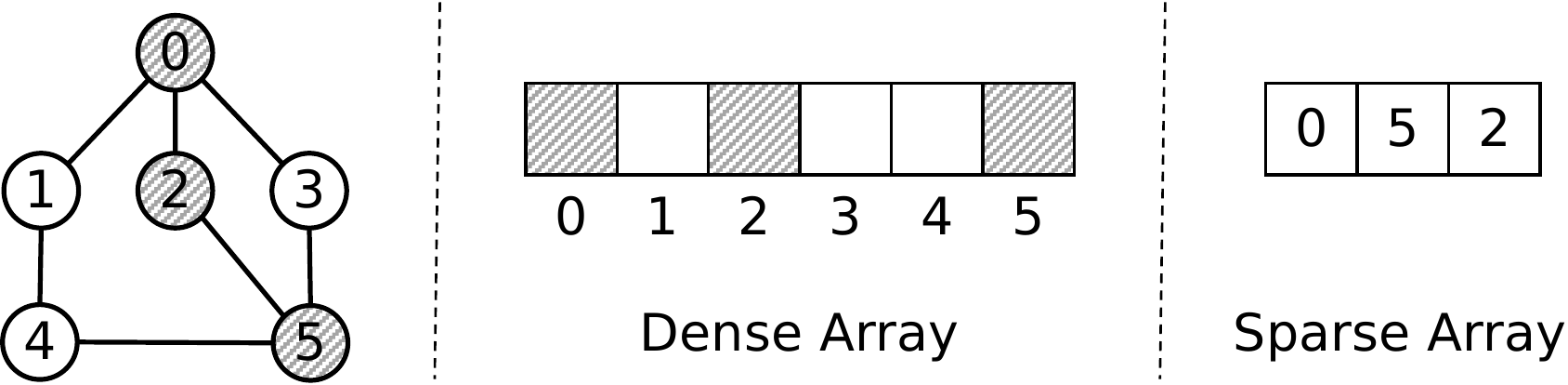}
\caption{Active $v_0$, $v_2$ and $v_5$ in \textit{dense array} and \textit{sparse array}}
\label{fig:dense_sparse_array}
\end{figure}

Recent graph processing systems~\cite{zhu_gemini_2016,zhangyunming_graphit_2018} and graph stores~\cite{kumar_graphone:_2019,zhu2019livegraph} prefer \textit{dense arrays} or \textit{bitmaps} to store active vertices because dense representations perform better than \textit{sparse arrays} in graph computing. Figure~\ref{fig:dense_sparse_array} shows an example of \textit{dense array} and \textit{sparse array}. KickStarter also uses \textit{bitmaps}, however, checking the entire vertex set and clearing the \textit{bitmaps} are expensive for incremental computing. For example, clearing and checking \textit{bitmaps} take KickStarter 90.3\% of the BFS computation time on Twitter-2010.

For per-update analysis and incremental computing, \emph{sparse arrays} can avoid accessing unnecessary vertices and reduce the average computing time from more than 50 ms to a few microseconds. In other cases, they can still improve computing performance when the batch-size is less than 200K on Twitter-2010. The reason is that \emph{sparse arrays} are not optimal, but still acceptable for many active vertices or even the entire set~\cite{shun_ligra_2013}. For example, to compute BFS on Twitter-2010 directly instead of incrementally, it takes \SYS 2.21 s, while it takes GraphOne 0.76 s with \emph{dense arrays}.

In summary, \emph{sparse arrays} provide orders of magnitude improvement for per-update analysis and handle most batch sizes well. Meanwhile, \emph{sparse arrays} are also acceptable for corner cases or even whole-graph analysis. Taking BFS on Twitter-2010 as an example, \SYS's performance only drops by 26.6\% with batches of 200M edges and drops by 65.6\% when re-computing BFS.
Therefore, we choose \emph{sparse arrays} to store active vertices.

We also manage to localize data access in other parts of the computation. For example, incremental computing relies on the previous version of results. In our implementation, we use \emph{sparse arrays} to track updates on results, while KickStarter copies the entire vertex set for every new iteration of analysis.

\begin{figure}[t]
\centering
\includegraphics[width=0.85\columnwidth]{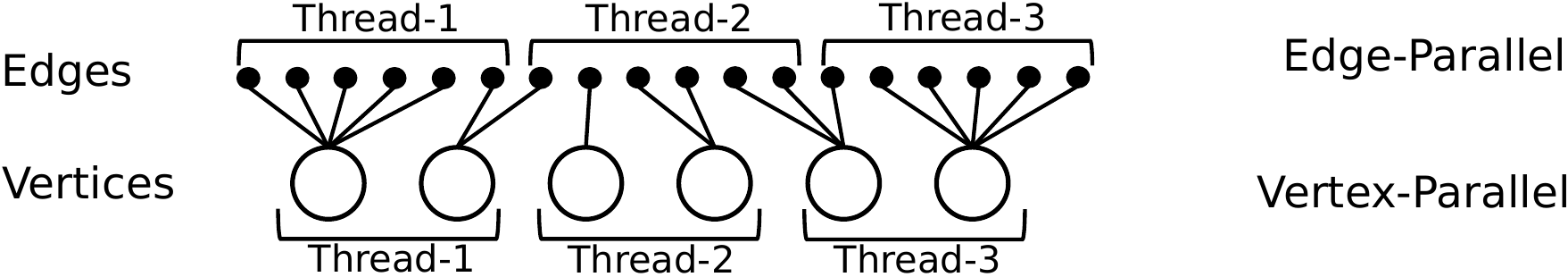}
\caption{\textit{Edge-parallel} and \textit{vertex-parallel} with three threads}
\label{fig:vertex_edge_parallel}
\end{figure}

\begin{figure}[t]
\centering
\includegraphics[width=0.75\columnwidth]{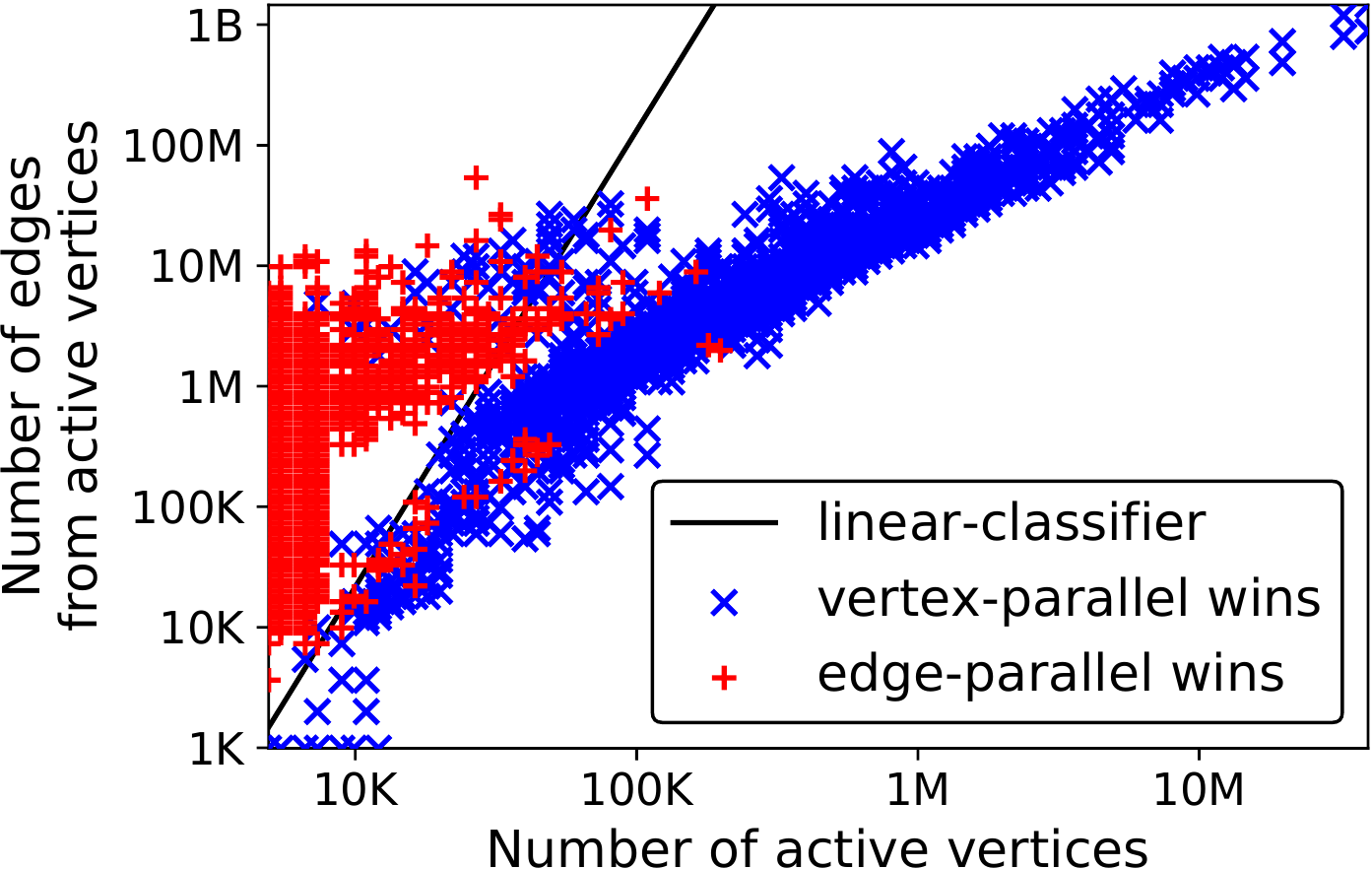}
\caption{Comparison of \textit{edge-parallel} and \textit{vertex-parallel}}
\label{fig:edge-vertex-stat}
\end{figure}

\paragraph{\bfseries Hybrid Parallel Mode.}

With sparse arrays, there are two parallel modes for push operations, \textit{vertex-parallel} and \textit{edge-parallel}. \textit{Vertex-parallel} takes active vertices as parallel units, while \textit{edge-parallel} is fine-grained that parallelizes across all edges instead of just vertices, as shown in Figure~\ref{fig:vertex_edge_parallel}. A common conclusion in graph computing is that \textit{vertex-parallel} is always better than \textit{edge-parallel}~\cite{zhangyunming_graphit_2018}. However, \textit{edge-parallel} sometimes 
outperforms \textit{vertex-parallel} in incremental computing, especially considering skewed distributions of degrees in power-law graphs. \textit{Edge-parallel} loses some locality of edges, but can provide more parallelism and better load balance when the number of active vertices is small.

Figure~\ref{fig:edge-vertex-stat} shows the results of comparing \textit{edge-parallel} and \textit{vertex-parallel} on the UK-2007~\cite{BoVWFI,BRSLLP} dataset running four different algorithms (BFS, SSSP, SSWP and WCC). The x-axis is the number of active vertices, and the y-axis is the out-degrees of active vertices. We average the time of push operations, and only keep the results where the difference is more significant than 20\% (filtering out 32\% results). Red dots indicate where \textit{edge-parallel} outperforms, and blue crosses show that where \textit{vertex-parallel} wins. When there are fewer active vertices and more active edges (top left corner of the figure), \textit{edge-parallel} is better than \textit{vertex-parallel}. 

We integrate \textit{edge-parallel} and \textit{vertex-parallel} by a linear classifier (the black straight line in Figure~\ref{fig:edge-vertex-stat}), which is trained by linear regression.
In our evaluations, the hybrid mode outperforms 24.2\% than the commonly used \emph{vertex-parallel} only mode. 


\section{Inter-update Parallelism}
\label{sec:opt_throughput}
\SYS has significantly cut the average processing time of per-update analysis by localized data access to about one-thousandth of the existing systems, which naturally turn into about 1000$\times$ throughput improvement (see Section~\ref{subsec:comparisons} for evaluation). 
However, the current design of single writer mechanism processes updates one by one under per-update analysis and limits \SYS's throughput because incremental computing depends on the previous results.
To further improve the throughput, we investigate how to process updates in parallel among multiple clients or user sessions while preserving the correctness and analysis frequency. 


A general approach is to employ transaction techniques from databases for parallel updates and analysis. Concurrency control mechanisms provide ACID properties and guarantee the correctness by \emph{serializable isolation}. Nevertheless, general read/write-set based concurrency control mechanisms are not practical in our scenario. When processing power-law graphs, the sizes of read/write sets are much larger than typical transaction sizes in online transaction processing (OLTP) databases. For example, the average size of read/write sets exceeds one hundred and sometimes reaches millions, when incrementally analyzing shortest paths on UK-2007.

\paragraph{\bfseries Observation.}


\begin{table}[t]
\caption{Datasets used in the experiments}
\vspace{-1em}
\label{tab:datasets}
\center
\resizebox{1.0\columnwidth}{!}{
\setlength{\tabcolsep}{0.2em}
\begin{tabular}{l|ccccccc}
\hline
Graph Dataset                             & Abbr. & Vertices & Edges & Temporal & Type & Root & Visited \\ \hline
HepPh~\cite{network_repository}           & PH   & 281K     & 4.60M & \checkmark & Collab. & 1 & 98\% \\
Wiki~\cite{network_repository}            & WK   & 2.13M    & 9.00M & \checkmark & Int. & 0 & 89\% \\
Flickr~\cite{network_repository}          & FC   & 2.30M    & 33.1M & \checkmark & Social & 1 & 82\% \\
StackOverflow~\cite{snapnets}             & SO   & 2.60M    & 63.5M & \checkmark & Int. & 0 & 78\% \\
BitCoin~\cite{network_repository}         & BC   & 24.6M    & 123M & \checkmark & Txn. & 2 & 49\% \\
SNB-SF-1000~\cite{erling_ldbc_2015}       & SB   & 3.14M    & 202M & \checkmark & Social & 0 & 84\% \\
LinkBench~\cite{armstrong_linkbench:_2013}  & LB & 128M    & 560M & \checkmark & Social  & 0 & 26\% \\
Twitter-2010~\cite{snapnets}              & TT   & 41.7M    & 1.47B &          & Social & 0 & 83\% \\
Subdomain~\cite{noauthor_wdc_nodate}      & SD   & 102M     & 2.04B &          & Web & 0 & 67\% \\
UK-2007~\cite{BoVWFI,BRSLLP}              & UK   & 106M     & 3.74B &          & Web & 0 & 91\% \\ \hline
\end{tabular}
}
\vspace{1em}
\caption{The proportion of updates which modify the results}
\vspace{-1em}
\label{tab:unsafe_proportion}
\center
\resizebox{\columnwidth}{!}{
\setlength{\tabcolsep}{0.25em}
\centering
\begin{tabular}{l|lcr|lcr|lcr|lcr} 
\hline
\multicolumn{1}{c|}{} & \multicolumn{3}{c|}{BFS} & \multicolumn{3}{c|}{SSSP} & \multicolumn{3}{c|}{SSWP} & \multicolumn{3}{c}{WCC}  \\ 
\hline
                      & 10\% & 50\% & 90\%       & 10\% & 50\% & 90\%        & 10\% & 50\% & 90\%        & 10\% & 50\% & 90\%       \\ 
\hline

PH                    & 0.03 & 0.03 & 0.01       & 0.03 & 0.03 & 0.01        & 0.02 & 0.04 & 0.03        & 0.04 & 0.02 & 0.01       \\
WK                    & 0.07 & 0.06 & 0.11       & 0.07 & 0.05 & 0.13        & 0.07 & 0.04 & 0.12        & 0.13 & 0.10 & 0.10       \\
FC                    & 0.00 & 0.01 & 0.06       & 0.00 & 0.01 & 0.06        & 0.00 & 0.01 & 0.05        & 0.11 & 0.04 & 0.03       \\
SO                    & 0.07 & 0.06 & 0.06       & 0.07 & 0.06 & 0.06        & 0.07 & 0.06 & 0.06        & 0.08 & 0.04 & 0.05       \\
BC                    & 0.00 & 0.10 & 0.15       & 0.00 & 0.11 & 0.18        & 0.00 & 0.09 & 0.15        & 0.45 & 0.47 & 0.50       \\
SB                    & 0.01 & 0.01 & 0.01       & 0.02 & 0.02 & 0.01        & 0.01 & 0.01 & 0.02        & 0.11 & 0.04 & 0.03       \\
LB                    & 0.00 & 0.03 & 0.09       & 0.00 & 0.03 & 0.09        & 0.00 & 0.03 & 0.09        & 0.22 & 0.16 & 0.12       \\
TT                    & 0.00 & 0.03 & 0.02       & 0.00 & 0.04 & 0.02        & 0.00 & 0.03 & 0.02        & 0.18 & 0.03 & 0.01       \\
SD                    & 0.00 & 0.04 & 0.03       & 0.00 & 0.04 & 0.03        & 0.00 & 0.03 & 0.03        & 0.23 & 0.05 & 0.03       \\
UK                    & 0.00 & 0.02 & 0.02       & 0.00 & 0.02 & 0.02        & 0.00 & 0.02 & 0.02        & 0.19 & 0.03 & 0.02       \\
\hline
\end{tabular}
}
\vspace{-1em}
\end{table}

According to the incremental computing model, the results of all vertices only depend on the edges on the dependency tree. Since each vertex has at most one parent in the dependency tree, there are at most $\left|V\right|$ (number of vertices) edges on the tree, rather than $\left|E\right|$ (total edges in the graph). The other edges are irrelevant to the results, therefore, the modification of these edges will not change any result. 


This fact inspires us and leads to a natural question, how much the updates will change the results.
We analyze four algorithms (BFS, SSSP, SSWP and WCC) on ten graphs listed in Table~\ref{tab:datasets}. We also vary the number of initially loaded edges ($\left|E\right|$), including 10\%, 50\% and 90\% of edges, to show the effect of the average degree ($\left|E\right| \mathbin{/} \left|V\right|$). The method of generating updates and varying degrees is the same as the evaluation in Section~\ref{sec:evaluation}. Table~\ref{tab:datasets} also enumerates the roots selection and the percentage of visited vertices from the root for BFS, SSSP and SSWP with 90\% edges. 

We find that only a small part of updates change the results for most cases, as shown in Table~\ref{tab:unsafe_proportion}. The proportion of updates which modify the results is less than 20\% in 115 combinations of algorithms and datasets (120 experiments in total). In 100/120 experiments, the proportion is less than 10\%. 77/120 experiments show that less than 5\% updates modify results. 

The observation guides us to propose a domain specific concurrency control mechanism for monotonic algorithms to avoid tracing memory accesses.
We name the updates which do not change any results as \emph{safe} updates. Correspondingly, \emph{unsafe} updates modify results or the dependency tree. If we identify \emph{safe} updates and only process them in parallel, not only can we preserve the correctness of per-update analysis, but also improve the throughput. 

\paragraph{\bfseries Classification of Updates.}

\begin{figure}[t]
\centering
\includegraphics[width=0.9\columnwidth]{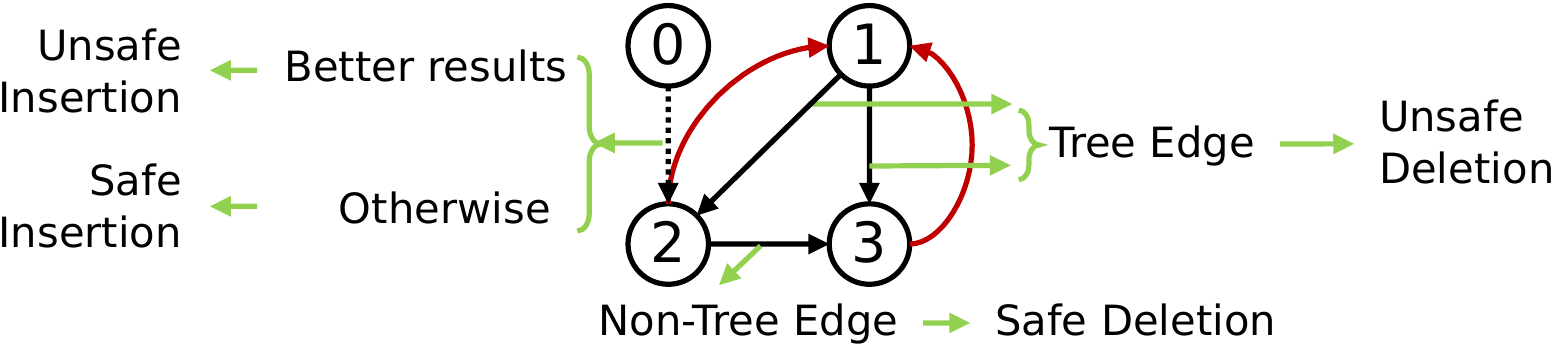}
\caption{\emph{Safe} updates and \emph{unsafe} updates}
\label{fig:safe_unsafe}
\vspace{-1em}
\end{figure}

Figure~\ref{fig:safe_unsafe} shows an example of \textit{safe} updates and \textit{unsafe} updates. 
Dark red arc arrows in the figure represent the dependency tree of the selected monotonic algorithm.
Formally, given a directed graph $G=(V,E)$ and the current dependency tree $T=(V_T,E_T)$, an update to the graph is considered \emph{safe} if and only if it fits in the following categories:

    \noindent \textbf{(1)} \texttt{\textcolor{Blue}{\textbf{\lstinline|ins_vertex|}}(v)} or \texttt{\textcolor{Blue}{\textbf{\lstinline|del_vertex|}}(v)} for any vertex $v$. The operation is valid only when $v$ is a isolated vertex (users must first delete all edges related to $v$ before deleting $v$), thus will not affect the results in monotonic algorithms~\cite{vora_kickstarter:_2017}.
    
    \noindent \textbf{(2)} \texttt{\textcolor{Blue}{\textbf{\lstinline|del_edge|}}(e)} for $e \notin E_T$, such as the edge $\langle v_2, v_3 \rangle$. These deletions will not modify the dependency tree, and therefore will not change any result. In contrast, deleting $\langle v_1, v_2 \rangle$ or $\langle v_1, v_3 \rangle$ will invalidate the states of $v_2$ or $v_3$ respectively.
    
    \noindent \textbf{(3)} \texttt{\textcolor{Blue}{\textbf{\lstinline|ins_edge|}}(e)} for $e = \langle v_s, v_t \rangle$ when \texttt{\textcolor{Blue}{\textbf{\lstinline|need_upd|}}(vt, vt.data, \textcolor{Blue}{\textbf{\lstinline|gen_next|}}(e, vs.data))} is \texttt{\lstinline|false|}. Taking edge insertion $\langle v_0, v_2 \rangle$ as an example, we first compute the new result of the destination $v_2$ from the source $v_0$ and the new edge $e$. The insertion is \textit{safe} if $e$ cannot produce a better result than the current result of $v_2$.


All other types of updates are \emph{unsafe}. In summary, the classification only depends on updating edges, sources and destinations.
The classification of updates is light-weight in \SYS because it does not require any scanning.




\paragraph{\bfseries Epoch Loop Schema and Inter-update Parallelism.}

\begin{figure}[t]
\centering
\includegraphics[width=0.9\columnwidth]{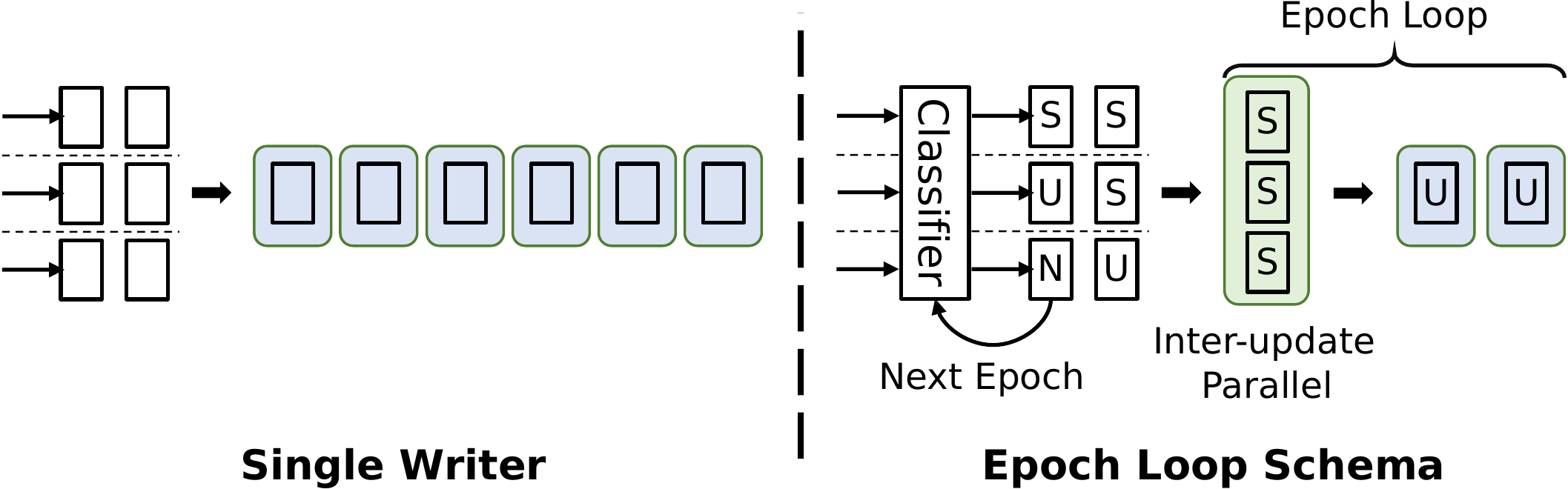}
\caption{\SYS's epoch loop schema}
\label{fig:epoch_loop}
\vspace{-1em}
\end{figure}


Then we design an epoch loop schema, which takes advantage of the parallelism brought by classification to improve throughput and ensure correctness. In each epoch, \SYS processes all \textit{safe} updates in parallel first and then handles all \textit{unsafe} updates one by one.
Figure~\ref{fig:epoch_loop} shows \SYS's epoch loop schema when processing updates from multiple sessions. There are six updates from three asynchronous sessions in the figure. \SYS classifies the updates into \textit{safe} (S), \textit{unsafe} (U), and \textit{next-epoch} (N).
After an \textit{unsafe} update, all updates in the same session are \textit{next-epoch} updates, which means they should be re-classified in the next epoch because any \textit{unsafe} operation could modify the results and change classifications of updates behind it. 
\SYS processes multiple \textit{safe} updates in parallel, exploiting inter-update parallelism. In contrast, \SYS handles \textit{unsafe} updates one by one and performs parallel incremental computing (intra-update parallelism) for each update.

For better user-friendliness, \SYS guarantees updates from a session will be executed by the same order of the updates and provides \emph{sequential consistency}.
However, it may lead to starvation. If some sessions keep producing \textit{safe} updates, \SYS will make \textit{unsafe} updates starved. In order to solve the starvation and also provide predictable processing time, we design a scheduler for \SYS. \SYS's scheduler controls the size of each epoch, and try to satisfy the user's desired tail latency (processing-time latency~\cite{karimov_benchmarking_2018}) as much as possible, which is discussed in Section~\ref{subsec:scheduler}.

In our experiments (Section~\ref{subsec:performance}), \SYS's throughput is 14.1$\times$ better than the throughput without the inter-update parallelism, while the 99.9th percentile (P999) latency is under 20 ms. The results show that the epoch loop schema, inter-update parallelism, and scheduler can effectively optimize \SYS's throughput.

\paragraph{\bfseries Supporting Transactions and Multiple Algorithms.}

We have shown that 
\SYS can support the situation where all updates are single vertex or edge updates, and only one algorithm is online in the system at a time.
Sometimes, users also expect the support of transactions or atomic batches containing multiple updates and also running multiple algorithms at the same time.
To support write transactions, we classify and process updates of a transaction as a whole. A write-only transaction is \textit{safe} only when all of its write operations are \textit{safe}, such as a transaction consisting of inserting a vertex $v_4$ and deleting the edge $\langle v_2, v_3 \rangle$ in Figure~\ref{fig:safe_unsafe}.
As for read-write transactions, although they are not typical for streaming systems, \SYS can still support them by treating them as \textit{unsafe} transactions and processing them individually by blocking other sessions (just long-term \textit{unsafe} updates in the epoch loops). 

Similar to write-only transactions, when maintaining multiple algorithms simultaneously, an update is \textit{safe} only when it is \textit{safe} for every algorithm. The proportion of \textit{safe} updates would decrease with larger transactions or more algorithms, which reduces the throughput brought from the inter-update parallelism. Compared with updates without being packed into transactions, \SYS's throughput reduces by 51.1\% on average when the size of each transaction is 16. 
In any case, even if all transactions or updates are \emph{unsafe}, \emph{localized data access} can still provide thousands of times throughput of existing systems.


\section{Implementation}
\label{sec:design}

\paragraph{\bfseries Graph Store.}
\label{subsubsec:graph_storage}
\SYS proposes \emph{Indexed Adjacency Lists}. 
Each vertex maintains its outgoing edges in an dynamic array (doubling capacity when full). Adjacency lists store directed edges, consist of the destination vertex IDs, the weight of each edge and the number of duplicated edges (the destination and the weight are both the same). 
\SYS also stores a transpose graph required by the incremental model.
Each vertex, whose degree is greater than a threshold, also contains an index, which represents the location of the edge in the list. The key of an edge is a pair of its destination vertex ID and its weight. The threshold provides a trade-off between memory consumption and lookup performance. We search it in the power of two to maximize performance divided by the square root of the memory usage (more performance-oriented). In our implementations, the threshold is 512. Users can search a better threshold based on their data, hardware, and requirements.

When inserting an edge, \SYS first checks whether the edge exists in adjacency lists from edge indexes. If the edge exists, \SYS only modifies the number of duplicated edges; otherwise \SYS appends the new edge to the adjacency list and updates the index. For deleting an edge, \SYS modifies the number of edges after searching from indexes. \SYS keeps tomb (deleted) edges first, and recycle them and their indexes when doubling the adjacency list. 
\SYS recycles the vertex IDs of deleted vertices into a pool.
When inserting vertex, \SYS either uses an ID from the recycling pool or assigns a new vertex ID.

\SYS uses Hash Table\footnote{Implemented by Google Dense Hashmap (\url{https://bit.ly/3rWs6yr}) and MurmurHash3} as the default indexes to obtain the average $\text{O}(1)$ time complexity of insertions and deletions.
There are also many alternative data structures that can replace Hash Table for indexes, such as BTree and ARTree~\cite{leis_adaptive_2013} (the adaptive radix tree). According to the theoretical complexity and our experiments, the performance of Hash Table is the best in general.
To maximize performance, we choose Hash Table by default although it is non-optimal in memory consumption. \SYS can also utilize other data structures for indexes if the memory capacity is a constraint.


\paragraph{\bfseries Tree and Value Store.}
\label{subsubsec:tree_and_value_storage}
\SYS stores dependency trees by parent pointer trees~\cite{parent_pointer_tree_2019}. Similar to KickStarter, each vertex maintains at most one \textit{bottom-up} pointer to its parent on the dependency tree. It is efficient to classify updates by checking whether the updating edge is a \textit{bottom-up} pointer on the dependency tree with parent pointer trees. 
During computing, modifications on parent pointer trees are also more lightweight than \textit{top-down} pointer trees. With \textit{top-down} pointers, updating the value of a vertex requires locking three vertices. On the contrary, parent pointer trees lock or atomically update the modified vertex only once. 
\paragraph{\bfseries History Store.}
\label{subsubsec:history_storage}
The history store consists of a doubly-linked list from new versions to old versions for each vertex, and sparse arrays for each version to trace modifications of the results. Linked lists are similar to version chaining in multi-versioned databases, which are generally efficient in practice~\cite{wu_empirical_2017}.

The history store only maintains short-term historical information, to provide consistent snapshots of the results. Every second, \SYS chooses the latest useless version among versions released by every session, marks the version and the previous versions as garbage, and then aggressively recycles them from sparse arrays.
For linked lists, \SYS performs lazy garbage collection, which means \SYS only recycles garbage from the tail of a vertex's linked list when a new version updates this vertex.

\paragraph{\bfseries Graph Computing Engine.}
\label{subsec:computing_engine}

\SYS chooses \textit{sparse arrays} to store active vertices and converts them to \textit{bitmaps} only when performing \textit{pull} operations (checking all incoming edges for each vertex). We create a separate \textit{sparse array} for each thread, which helps eliminating the overhead of synchronization and contention among multiple threads.


For \textit{push} operations, we propose a hybrid mode supported by a linear regression based classifier to adaptively choose the proper \textit{edge-parallel} or \textit{vertex-parallel} mode. 
In our implementation, we train the classifier based on UK-2007 dataset, and it works well on other graphs. In our evaluation, the hybrid mode reduces \SYS's computing time by 24.2\% compared to \textit{vertex-parallel} only mode, showing the effectiveness of the classifier. Users can also train the classifier by their datasets and algorithms. Online training would bring additional overhead, so we choose to fix the parameters first and leave online training as our future work.

\paragraph{\bfseries Scheduler.}
\label{subsec:scheduler}

The purpose of the scheduler in \SYS is to avoid starvation and also to achieve the highest possible throughput automatically under an expected tail latency. Avoidance of starvation has been discussed in Section~\ref{sec:opt_throughput}, so we focus on how the scheduler improves throughput while maintaining the latency demand.

The scheduler would first try to pack as many safe updates in an epoch-loop as possible to maximize throughput. It may break the latency constraint without a latency control, so the scheduler uses two heuristics to abort parallel execution of safe updates and turns to process unsafe updates. 
One is when the waiting time of the earliest \textit{unsafe} update in the queue almost exceeds the target latency. In our implementation, we set the target latency to 0.8 times the user-specified latency limitation. 
Another one is when the number of unprocessed \textit{unsafe} updates reached a dynamic threshold. \SYS adjusts the threshold based on historical information.
If the proportion of qualified updates (under the latency limitation) is higher than the target after the last adjustment, the scheduler will slowly increase the threshold. Otherwise, if the proportion is lower than the goal, the scheduler will quickly decrease the threshold. 
In our implementation, the initial threshold is the number of physical threads, and \SYS adjusts the threshold every three epoch-loops. \SYS increases the threshold by 1\% each time, and when decreasing, adjusts the threshold by 10\%.
Since the scheduler is self-adjusting, it can support various algorithms and datasets.

In our experiments, the scheduler can meet the latency requirement, and get a good trade-off between latency and throughput. 
\vspace{-1.5em}
\section{Evaluation}
\label{sec:evaluation}


\vspace{-0.1em}
\subsection{Experimental Setup}
\label{subsec:setup}

We evaluate \SYS by four algorithms implemented by Algorithm API in Table~\ref{tab:UDF_use_case}, including Breadth-First Search (BFS), Single Source Shortest Path (SSSP), Single Source Widest Path (SSWP) and Weakly Connected Component (WCC), on ten graph datasets in Table~\ref{tab:datasets}.
LinkBench (128M vertices) and LDBC SNB (SF1000) are interactive datasets from graph database benchmarks~\cite{armstrong_linkbench:_2013,erling_ldbc_2015}, consisting of a pre-populated graph and incoming updates. For the other datasets, we split the edge set similar to KickStarter, GraphBolt, and SAMS~\cite{then_automatic_2017}, by pre-populating a part of edges and treating the other edges as updates. 
We load 90\% edges first, select 10\% edges as the deletion updates from loaded edges, and treat the remaining (10\%) edges as the insertion updates. If datasets are timestamped, we choose the latest 10\% as the insertion set and the oldest 10\% as the deletion set; otherwise, we randomly select edges as updates. 
The ratio of insertions to deletions is 50\% by default, and we alternately request insertions and deletions of each edge. We also name the entire pre-loaded graphs in LinkBench and SNB as graphs with 90\% edges for alignment of the results.



We set up experiments on two dual-socket servers, running \SYS and clients respectively. Each server has two Intel Xeon Gold 6126 CPU (12 physical cores per CPU), 576GB main memory, an Intel Optane P4800X 750GB SSD, an 100Gb/s Infiniband NIC, and runs Ubuntu 18.04 with Linux 4.15 kernel.



\vspace{-0.7em}
\subsection{Performance of \SYS}
\label{subsec:performance}
We evaluate the performance of \SYS by a group of emulated synchronous users, similar to TPC-C~\cite{noauthor_tpc-c_nodate}. Remote clients interact with \SYS by RPCs. To eliminate the impact of the network, clients connect to \SYS by Infiniband network and a light-weight RPC framework with the RDMA technique. Each client maintains multiple sessions, which represent emulated users. The users repeatedly send a single update and wait for the response. 
Here we focus on edge insertions and edge deletions because they can express all types of updates regarding edges and vertices.


Latency (more precisely, processing-time latency~\cite{karimov_benchmarking_2018}) is measured on the client side and defined as the elapsed time between the request and response. We calculate throughput by the total execution time and the number of updates. The latency requirement is that at least 99.9\% of updates should receive responses in 20 milliseconds. Such a strict latency target requires the system to provide sufficient real-time ability. Updates in the evaluation are insertions and deletions with only a single edge, which evaluate the performance with the minimal granularity of updates and analysis.  

\begin{figure*}[ht]
\centering
\begin{minipage}{\textwidth}
	\centering
    \includegraphics[width=\columnwidth]{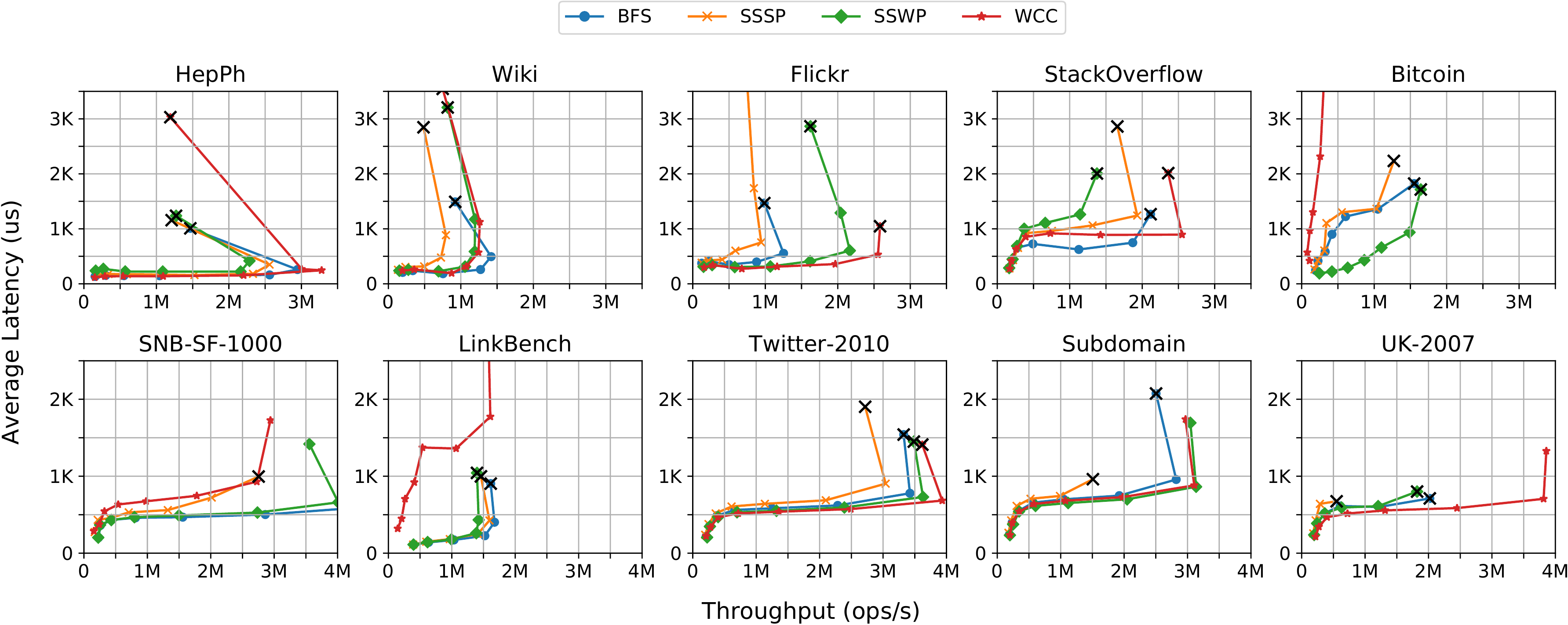}
    \subcaption{Performance trends of throughput and average latency in \SYS}
    \vspace{0.5em}
\end{minipage}
\begin{minipage}{\linewidth}
\center
\resizebox{\linewidth}{!}{
\setlength{\tabcolsep}{0.35em}
\begin{tabular}{r|ccc|ccc|ccc|ccc} 
\hline
   & \multicolumn{3}{c|}{BFS}        & \multicolumn{3}{c|}{SSSP}       & \multicolumn{3}{c|}{SSWP}       & \multicolumn{3}{c}{WCC}          \\ 
\hline
   & T. (op/s) & Mean (us) & P999 (ms) & T. (op/s) & Mean (us) & P999 (ms) & T. (op/s) & Mean (us) & P999 (ms) & T. (op/s) & Mean (us) & P999 (ms)  \\ 
\hline
HepPh (PH) & 2.95M     & 258.44    & 12.13    & 2.56M     & 348.62    & 18.59    & 2.28M     & 416.78    & 19.92    & 3.28M     & 244.63    & 6.234     \\
Wiki (WK) & 1.42M     & 492.96    & 10.34    & 794K      & 883.53    & 10.33    & 1.20M     & 1171.4    & 15.32    & 1.26M     & 1124.2    & 8.384     \\
Flickr (FC) & 1.25M     & 552.97    & 15.28    & 944K      & 752.21    & 10.34    & 2.17M     & 602.25    & 9.901    & 2.56M     & 530.92    & 7.584     \\
StackOverflow (SO) & 1.86M     & 747.43    & 19.89    & 1.93M     & 1240.7    & 19.76    & 1.14M     & 1260.6    & 18.75    & 2.55M     & 893.12    & 17.00     \\
BitCoin (BC) & 1.05M     & 1356.9    & 19.64    & 1.03M     & 1363.5    & 19.58    & 1.49M     & 935.56    & 16.52    & 432K      & 6614.2    & 16.72     \\
SNB-SF-1000 (SB) & 4.51M     & 601.72    & 14.43    & 2.02M     & 722.34    & 18.98    & 4.01M     & 660.22    & 12.37    & 2.94M     & 1726.5    & 18.84     \\
LinkBench (LB) & 1.67M     & 401.54    & 14.93    & 1.59M     & 425.11    & 18.85    & 1.42M     & 432.46    & 18.53    & 1.61M     & 1773.2    & 18.52     \\
Twitter-2010 (TT) & 3.42M     & 777.90    & 19.51    & 3.04M     & 903.07    & 18.37    & 3.63M     & 729.36    & 17.92    & 3.93M     & 682.75    & 18.13     \\
Subdomain (SD) & 2.82M     & 956.91    & 18.72    & 989K      & 742.90    & 18.71    & 3.14M     & 862.07    & 18.38    & 3.11M     & 881.11    & 18.50     \\
UK-2007 (UK) & 1.22M     & 600.17    & 18.22    & 288K      & 640.68    & 17.86    & 1.21M     & 610.28    & 18.68    & 3.86M     & 1327.4    & 18.14     \\
\hline
\end{tabular}
}
\subcaption{Metrics when \SYS reaches peak throughput}
\label{tab:peak_throughput}
\end{minipage}
\vspace{-1em}
\caption{\SYS's throughput and latency, while ensuring P999 latency within 20 milliseconds and per-update analysis}
\label{fig:throughput_latency}
\vspace{-0.5em}
\end{figure*}

All modules in \SYS are enabled, including the write-ahead log, the scheduler and the history store.
On the next page, Figure~\ref{fig:throughput_latency} indicates average latency and throughput when doubling the number of sessions from 48 (number of the hardware threads) until \SYS cannot satisfy the target latency, which is up to 6144 (64$ \times$48) sessions. Black crosses show where \SYS breaks the latency limitation. 
With more sessions, \SYS gets more opportunities to schedule and execute \textit{safe} updates in parallel for each epoch, so \SYS can get higher throughput. 

Figure~\ref{tab:peak_throughput} lists detailed metrics when throughput reaches the peak. 
In Figure~\ref{tab:peak_throughput}, T. represents throughput, while Mean and P999 are average latency and tail latency, respectively.
\SYS's throughput reaches hundreds of thousands or millions of updates per second, meanwhile, the P999 latency is under 20 milliseconds. These results indicate that the designs of \SYS can provide high throughput under per-update analysis.

\SYS's throughput is close to 100K ops/s without the inter-update parallelism, and can additionally improve by 15.5$\times$ for BFS, 9.93$\times$ for SSSP, 15.3$\times$ for SSWP, 17.1$\times$ for WCC, and 14.1$\times$ in overall when inter-update parallelism is enabled. 
Even if unsafe updates reach 49.7\% of the updates (WCC on Bitcoin), it can still provide 3.25$\times$ speedup. The improvements show that the epoch loop schema, inter-update parallelism, and scheduler can effectively optimize the throughput for per-update analysis.

\paragraph{\bfseries Performance under Different Configurations.}

Next, we adjust the number of pre-populated edges (representing the size of the sliding window), the proportion of insertions and deletions, and the size of the transactions to evaluate the robustness of the system. In this paragraph, we only present the geometric average of the peak throughput relative to the default configuration for each algorithm on the next page, due to the page limit. 

Our experiments keep the same number of edges as the pre-loaded size and perform sliding updates. The pre-loaded size can indicate the size of the sliding window. By default, we pre-populate 90\% of the graph. Table~\ref{tab:relateive_throughput_sliding_window} lists relative throughputs when the pre-populated part is 10\% and 50\% graph. For BFS, SSSP and SSWP, \SYS gets throughput benefit with 10\% and 50\% pre-loaded edges because fewer edges result in less visited vertices from the root. For WCC, the performance drops 15\% with 50\% edges and 66\% with 10\% edges because fewer edges build up sparser graph than 90\% edges, make the connected component unstable, generate more unsafe updates (see Table~\ref{tab:unsafe_proportion}) and require more computing.

Table~\ref{tab:relateive_throughput_add_ratio} shows the performance with varying percentage of edge insertions, compared to the performance with 50\% insertions. From the table, \SYS provides higher throughput as the proportion of insertions increases. The reason is that deletions need to reset results following to the dependency tree, while insertions do not.

We also evaluate \SYS using transactions of different sizes.
The latency constraint is that at least 99.9\% of updates should receive responses in 20 milliseconds. Each transaction contains a fixed number of updates. If the latency exceeds the transaction size multiplied by 20 milliseconds, the transaction is timeout. We still use the number of updates per second to express throughput. Table~\ref{tab:relateive_throughput_transaction} lists the results. When processing larger transactions, the throughput of \SYS will drop to a maximum of 61\% (WCC and 16 updates per transaction). The reason is that larger transactions lead to lower proportions of \textit{safe} transactions, which reduces the benefits from inter-updates parallel. Nevertheless, \SYS still supports several hundred thousands of updates per second.

\begin{table}[htb]
\caption{Relative throughput with different sliding window}
\vspace{-1em}
\label{tab:relateive_throughput_sliding_window}
\center
\resizebox{0.47\columnwidth}{!}{
\setlength{\tabcolsep}{0.4em}
	\centering
    \begin{tabular}{r|cccc}
    \hline
         & BFS  & SSSP & SSWP & WCC  \\ \hline
    50\% & 1.29 & 1.35 & 1.46 & 0.85 \\
    10\% & 2.23 & 3.29 & 2.26 & 0.34 \\ \hline
    \end{tabular}
}

\vspace{1em}
\caption{Relative throughput with varying insertion percent}
\vspace{-1em}
\label{tab:relateive_throughput_add_ratio}
\center
\resizebox{0.9\columnwidth}{!}{
\setlength{\tabcolsep}{0.3em}
	\centering
    \begin{tabular}{r|cccccr|cccc}
    \cline{1-5} \cline{7-11}
         & BFS  & SSSP & SSWP & WCC  &   &       & BFS  & SSSP & SSWP & WCC  \\ \cline{1-5} \cline{7-11} 
    0\%  & 0.72 & 0.79 & 0.88 & 0.67 &   & 75\%  & 1.09 & 1.01 & 1.04 & 1.10 \\
    25\% & 0.92 & 0.83 & 0.93 & 0.71 &   & 100\% & 1.20 & 1.08 & 1.15 & 1.34 \\ \cline{1-5} \cline{7-11} 
    \end{tabular}
}

\vspace{1em}

\caption{Relative throughput with varying transaction sizes}
\vspace{-1em}
\label{tab:relateive_throughput_transaction}
\center
\resizebox{0.9\columnwidth}{!}{
\setlength{\tabcolsep}{0.4em}
	\centering
    \begin{tabular}{r|cccccr|cccc}
    \cline{1-5} \cline{7-11}
      & BFS  & SSSP & SSWP & WCC  &   &    & BFS  & SSSP & SSWP & WCC  \\ \cline{1-5} \cline{7-11} 
    2 & 0.87 & 0.85 & 0.97 & 0.79 &   & 8  & 0.59 & 0.67 & 0.62 & 0.48 \\
    4 & 0.70 & 0.76 & 0.78 & 0.59 &   & 16 & 0.46 & 0.63 & 0.51 & 0.39 \\ \cline{1-5} \cline{7-11} 
    \end{tabular}
}

\vspace{-1em}
\end{table}

Finally, we evaluate \SYS when maintaining multiple algorithms simultaneously. \SYS maintains BFS, SSSP, SSWP and excludes WCC because these three algorithms require directed edges, but WCC requires undirected edges. We set the latency constraints to P999 and 60 milliseconds. The throughput is 1.20M ops/s for HepPh, 107K ops/s for Wiki, 391K ops/s for Flickr, 719K ops/s for StackOverflow, 429K ops/s for Bitcoin, 1.89M ops/s for LDBC-SNB, 288K ops/s for LinkBench, 1.61M ops/s for Twitter-2010, 891K ops/s for Subdomain and 363K ops/s for UK-2007.

The above results show that \SYS can handle various workloads robustly.

\begin{figure}[t]

\centering
\begin{minipage}{0.475\columnwidth}
	\centering
    \includegraphics[width=\columnwidth]{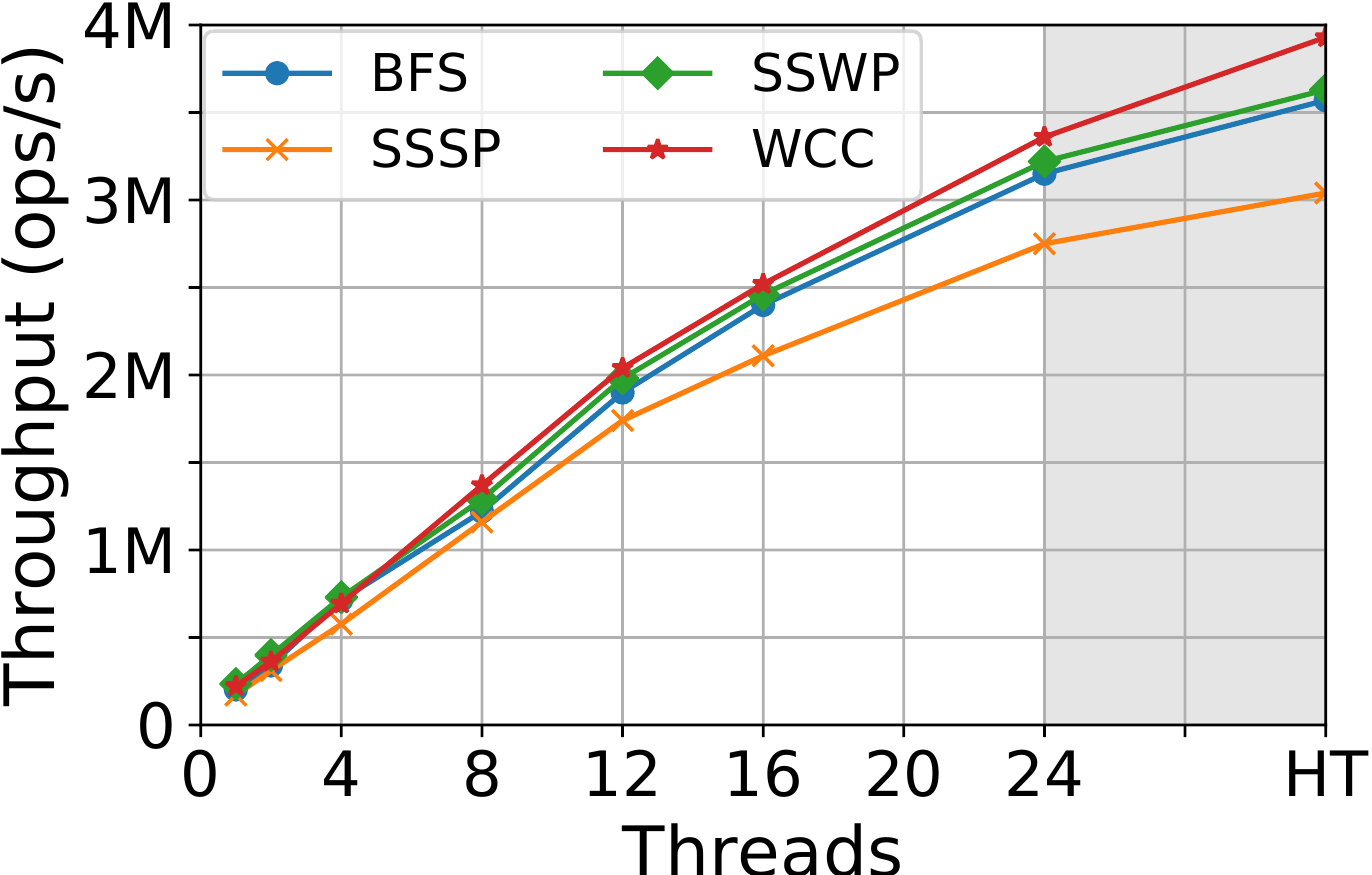}
    \subcaption{Scalability \label{fig:scalability}}
\end{minipage}
\begin{minipage}{0.505\columnwidth}
	\centering
    \includegraphics[width=\columnwidth]{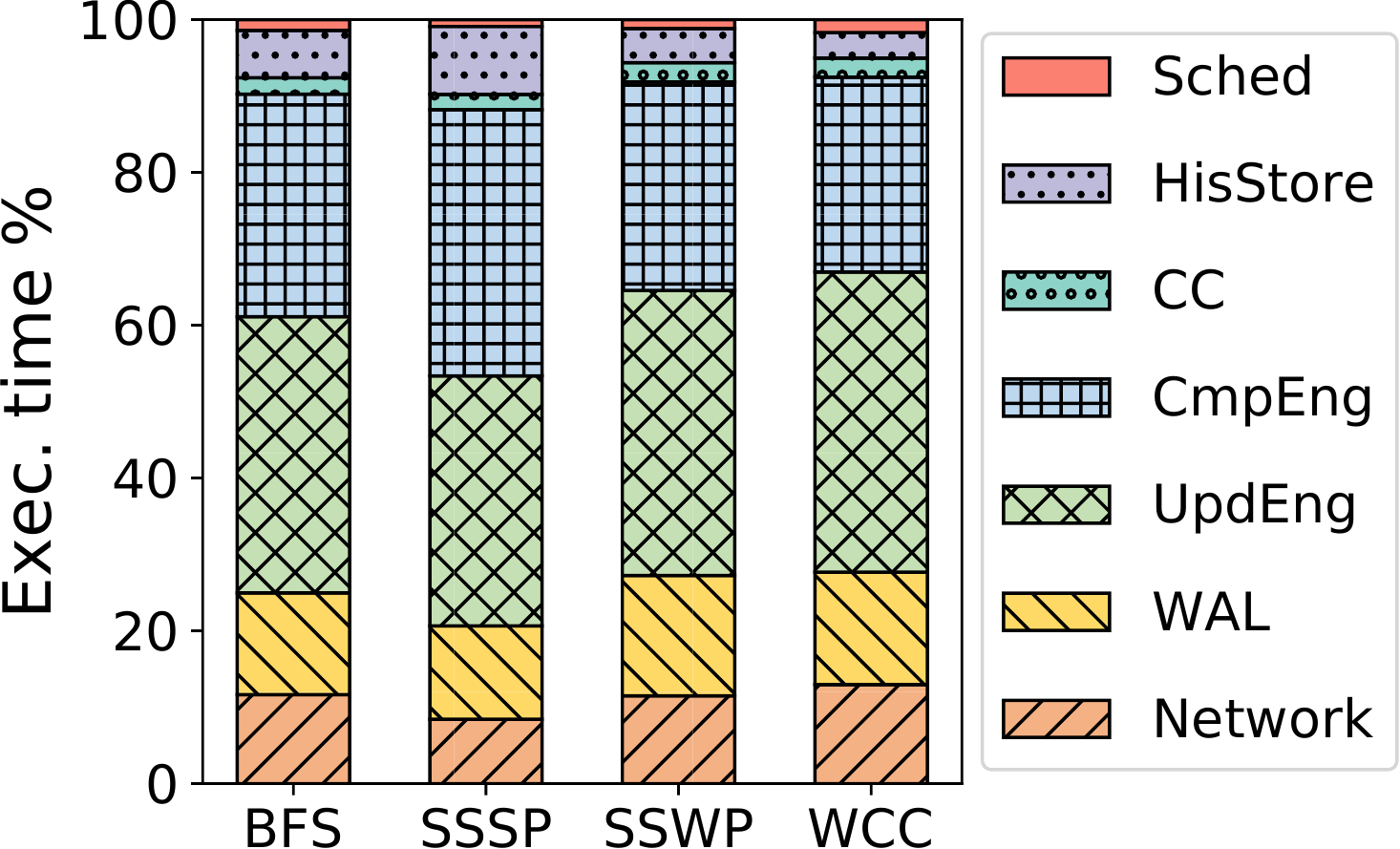}
    \subcaption{Performance Breakdown \label{fig:breakdown}}
\end{minipage}
\vspace{-1em}
\caption{Scalability and Performance Breakdown}
\label{fig:scalability_breakdown}
\vspace{-1em}
\end{figure}

\begin{figure}[t]
\centering
\includegraphics[width=0.8\columnwidth]{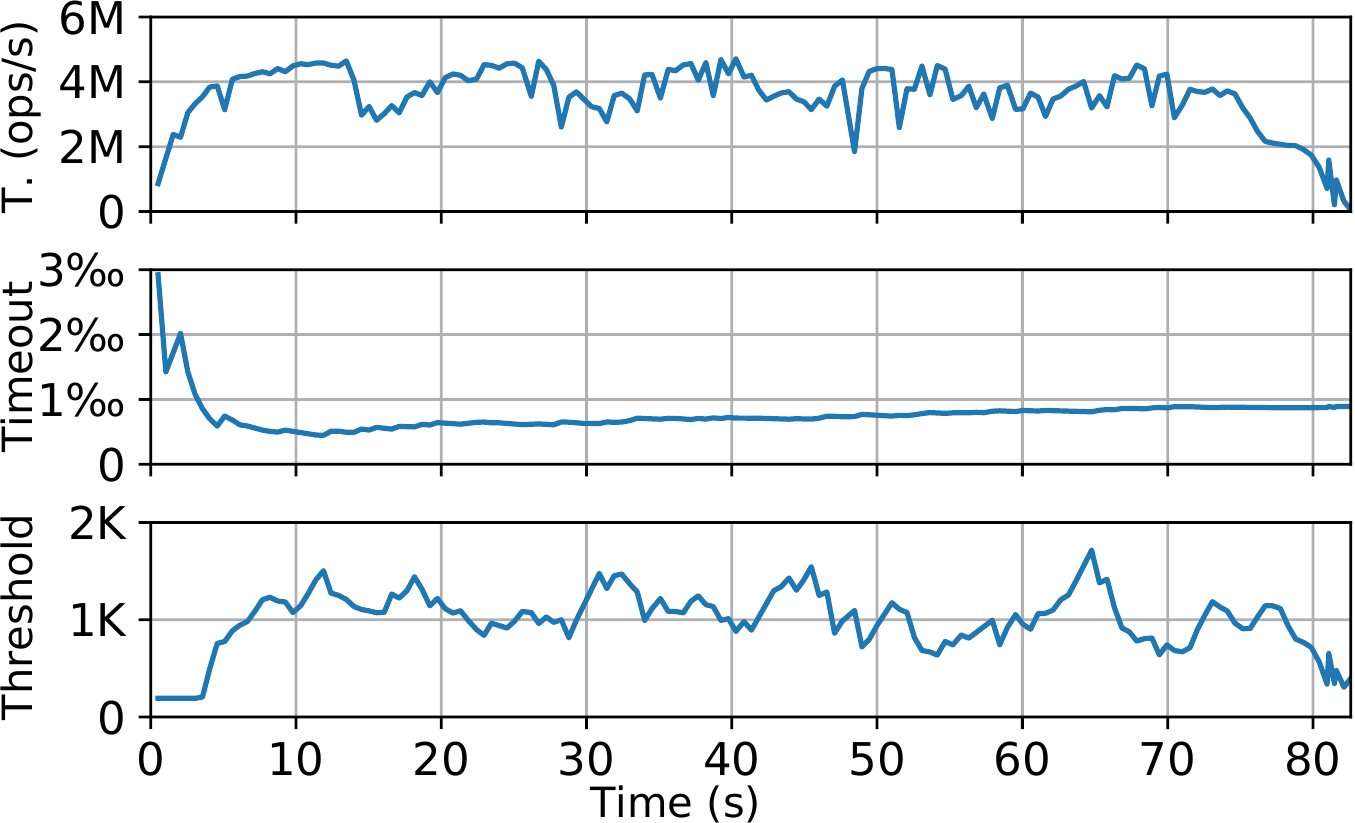}
\caption{Performance over time, sampling every 0.5 s}
\label{fig:scheduler}
\vspace{-0.5em}
\end{figure}

\paragraph{\bfseries Performance Case Study.}

We take Twitter-2010 dataset as an example and present more performance details for \SYS.

We first examine the multi-core scalability of \SYS under an increasing number of threads, as shown in Figure~\ref{fig:scalability}. \SYS's throughput scales smoothly with more cores until 24 physical cores are occupied, and further improves about 13.5\% with hyper threading. As a result, \SYS's throughput speedup is 17.6$\times$ for BFS, 17.8$\times$ for SSSP, 15.4$\times$ for SSWP and 17.7$\times$ for WCC with 24 physical cores (48 hyper threads).

We then look into \SYS's performance breakdown of components. Figure~\ref{fig:breakdown} illustrates that \SYS provides similar breakdowns under different algorithms. On an average of four algorithms, graph updating engine (UpdEng) and computing engine (CmpEng), as the core of \SYS, take 36.4\% and 29.2\% of the wall time respectively. To trace results, history store (HisStore) costs 5.7\% of the time. The concurrency control module (CC) and the scheduler (Sched) are very lightweight and only bring a total of 3.6\% overheads. WAL provides durability and network interacts with clients, which occupy 14.0\% and 11.1\% time, respectively.

We also trace \SYS's throughput (T.), timeout updates with more than 20 ms latency and the scheduler's threshold over time when maintaining BFS on Twitter-2010, as shown in Figure~\ref{fig:scheduler}. It shows that the scheduler can self-adjust its threshold to meet a tail-latency requirement and provide a good throughput over time.

\subsection{Comparison of Implementation Choices}
\label{subsec:eval_design}
To verify that our implementation choices are effective, we evaluate alternative choices by the algorithms and datasets used in Section~\ref{subsec:performance}. The scheduler and history store are disabled in this part. We classify updates first, apply all \textit{safe} updates in parallel, and then apply \textit{unsafe} updates one by one. The purpose is to separate \textit{safe} and \textit{unsafe} updates, and to show the impact of different designs.
\paragraph{\bfseries Graph Computing Engine.}
We first evaluate the graph computing engine by comparing it with the performance of \textit{vertex-parallel}, \textit{edge-parallel} and the \textit{hybrid-parallel} strategies. We focus on \textit{unsafe} updates and keep the adjacency lists stored by arrays to eliminate the impact of data structures.

\begin{figure}[t]

\centering
\begin{minipage}{0.40\columnwidth}
	\centering
    \includegraphics[width=\columnwidth]{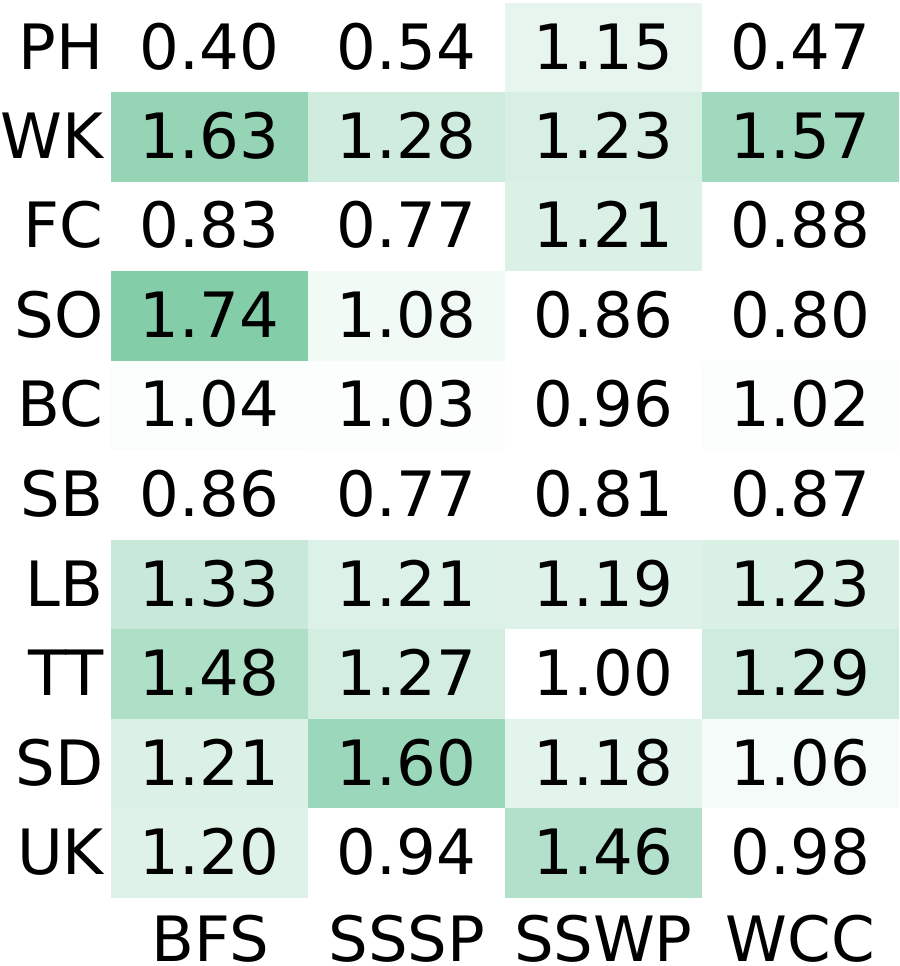}
    \subcaption{\textit{edge-parallel} \label{fig:computing_engine_edge}}
\end{minipage}
\hspace{0.04\columnwidth}
\begin{minipage}{0.40\columnwidth}
	\centering
    \includegraphics[width=\columnwidth]{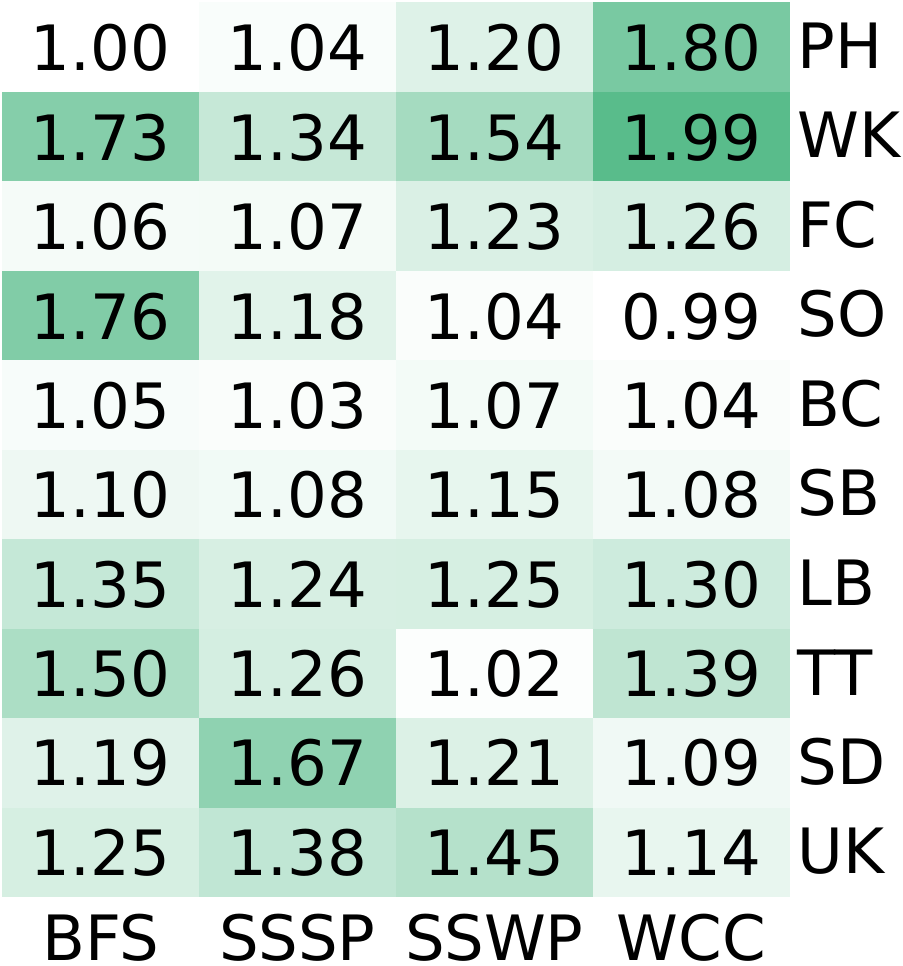}
    \subcaption{\textit{hybrid-parallel} \label{fig:computing_engine_hybrid}}
\end{minipage}
\vspace{-1em}
\caption{Speedup compared with \textit{vertex-parallel}}
\label{fig:computing_engine_speedup}
\vspace{-1em}
\end{figure}

Figure~\ref{fig:computing_engine_speedup} lists the speedup compared with \textit{vertex-parallel}. We measure the slowest 1\% updates first because the graph computing engine is the bottleneck for these updates, and they mainly affect \SYS's tail latency. According to Figure~\ref{fig:computing_engine_edge}, \textit{edge-parallel} is better than \textit{vertex-parallel} in some cases, which validates our discussion in Section~\ref{subsec:opt_computing_engine}. \textit{Hybrid-parallel} can integrate the advantages of \textit{vertex-parallel} and \textit{edge-parallel} well and provide better performance. It can accelerate computing up to 1.99 times, except for WCC on StackOverflow (a slight drop of 0.8\%).

The performance (geometrically averaged) of \textit{edge-parallel} outperforms \textit{vertex-parallel} by 3.9\%. The \textit{hybrid-parallel} strategy can achieve more improvements than \textit{edge-parallel}, reaching 1.24 times speedup over \textit{vertex-parallel} and 1.19 times over \textit{edge-parallel}. For all \textit{unsafe} updates, the performance advantages are 4.8\% and 6.1\%.




\paragraph{\bfseries Graph Store.}


We next evaluate six alternative data structures for the graph store. IA\_Suffix means the adjacency lists are stored in arrays and corresponding indexes. IO\_Suffix represents that \SYS only stores edges in the indexes. 
We evaluate three indexes, Dense Hash Table (Hash), BTree and ARTree.

Table~\ref{tab:data_structure_speedup} shows the relative overall performance measured from various data structures. The baseline is IA\_Hash used by \SYS.
We calculate the geometric average of the relative performance to reflect the overall performance. 

For \textit{safe} updates, IA\_Hash and IO\_Hash provide higher performance when processing graph updates because the time complexity of Hash Table is better than other indexes. IO\_Hash reduces the overhead by about 7\% compared to IA\_Hash because IO\_Hash does not maintain additional compact adjacency lists. \SYS pays overheads for adjacency lists to optimize the computing because \textit{unsafe} updates (with computing) mainly determine the tail latency of per-update analysis and take an average 2.59 times longer than \textit{safe} updates under large batches. The additional adjacency lists give IA\_Hash a 17\% advantage for \textit{unsafe} updates. 
Overall, \SYS's default data structure (IA\_Hash) performs well. 

\begin{table}[t]
\caption{Overall performance of data structures}
\vspace{-1em}
\label{tab:data_structure_speedup}
\center
\resizebox{0.90\columnwidth}{!}{
\begin{tabular}{r|ccc|ccc}
\hline
           & \multicolumn{3}{c|}{Index with Array (IA)} & \multicolumn{3}{c}{Index Only (IO)} \\ \cline{2-7} 
           & ARTree         & BTree         & Hash       & ARTree      & BTree       & Hash     \\ \hline
Safe    & 0.91    & 0.79   & 1.00     & 0.94    & 0.81   & 1.07     \\
Unsafe  & 0.93    & 0.96   & 1.00     & 0.48    & 0.76   & 0.83     \\ 
Overall & 0.92	  & 0.90   & 1.00	  & 0.57	& 0.78	 & 0.89     \\ \hline
\end{tabular}
}

\vspace{-0.5em}
\end{table}

\paragraph{\bfseries Memory Consumption.}

\begin{table}[t]
\caption{\SYS's memory usage relative to raw-data}
\vspace{-1em}
\label{tab:memory_consumption}
\center
\resizebox{0.92\columnwidth}{!}{
\setlength{\tabcolsep}{0.4em}
\begin{tabular}{r|ccc|ccc}
\hline
           & \multicolumn{3}{c|}{Index with Array (IA)} & \multicolumn{3}{c}{Index Only (IO)} \\ \cline{2-7} 
           & ARTree         & BTree         & Hash       & ARTree      & BTree       & Hash     \\ \hline
Unweighted & 3.63        & 2.36       & 3.25       & 3.45     & 2.10     & 2.97     \\
8B\_Weight & 3.45        & 2.50       & 3.38       & 3.13     & 2.17     & 3.04     \\ \hline
\end{tabular}
}

\vspace{-1em}
\end{table}

Table~\ref{tab:memory_consumption} shows the memory consumption of \SYS. We compare \SYS's memory footprint with raw-data (16 Bytes per edge for Unweighted graphs and 24 Bytes per edge for 8B\_Weight graphs) and geometrically average them. \SYS spends 3.25$\times$ memory on unweighted graphs and 3.38$\times$ memory on weighted graphs.
\SYS's indexes brings most of the memory overhead, but they are necessary to support both fast insertions and deletions, which are discussed in Section~\ref{subsec:opt_graph_storage}. 
In order to support efficient bi-direct traversal at the same time, \SYS maintains a transpose (reverse) of the directed graph, which doubles the memory occupation.
The adjacency lists only occupy less than 0.5$\times$ of raw-data memory according to the comparison of IA\_Hash with IO\_Hash because they only store the destination vertices. Meanwhile, \SYS only creates indexes for vertices whose degree exceeds the threshold to balance the memory consumption and performance.
If a compact memory footprint is necessary, it is a wise choice to replace Hash Table (IA\_Hash) with BTree (IA\_BTree), which can reduce memory usage by about 1.15 times raw-data and lose 22\% performance at the same time. 

Since \SYS is an in-memory system, we also explore how to scale for larger datasets. We try to extend \SYS to support out-of-core processing. We use \textit{mmap} to build a prototype that swaps to an SSD (Intel P3608 4TB SSD). We choose IA\_BTree as the data structure and run UK-2014~\cite{BoVWFI,BRSLLP} (788M vertices, 47.6B edges, 710GB raw data). For WCC, it can process 262K \textit{safe} updates per second. The average time of \textit{unsafe} updates is 147 us, and the P999 latency is 2091 us, showing that scaling up to disks is a feasible solution. We take scaling up and scaling out as our future work.

\begin{figure*}[htb]
\centering
\begin{minipage}{0.245\linewidth}
	\centering
    \includegraphics[width=\columnwidth]{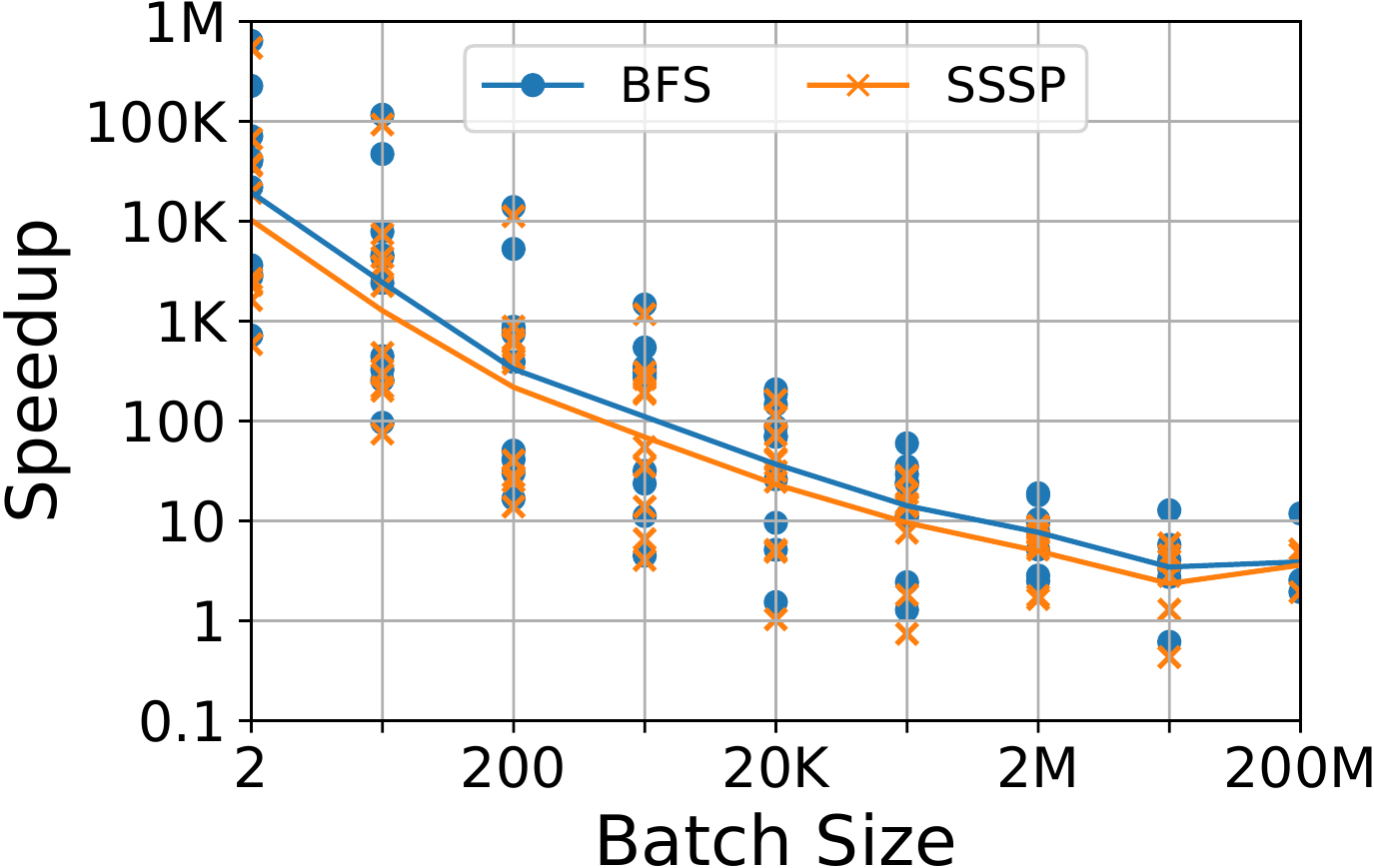}
    \subcaption{Speedup over KickStarter
    \label{fig:compare_kickstarter_speedup}}
\end{minipage}
\begin{minipage}{0.245\linewidth}
	\centering
    \includegraphics[width=\columnwidth]{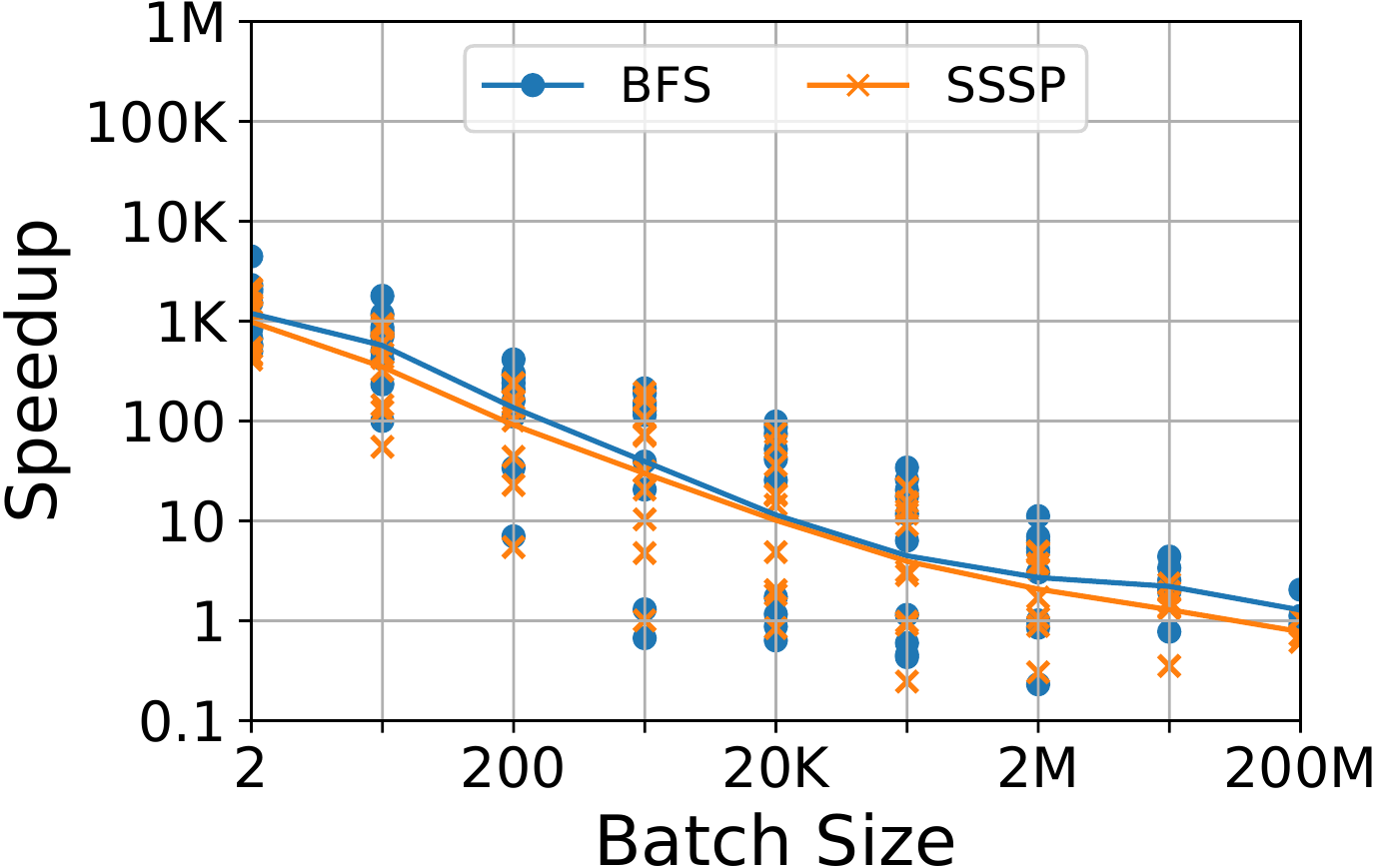}
    \subcaption{Speedup over DD.
    \label{fig:compare_differential_dataflow_speedup}}
\end{minipage}
\begin{minipage}{0.245\linewidth}
	\centering
    \includegraphics[width=\columnwidth]{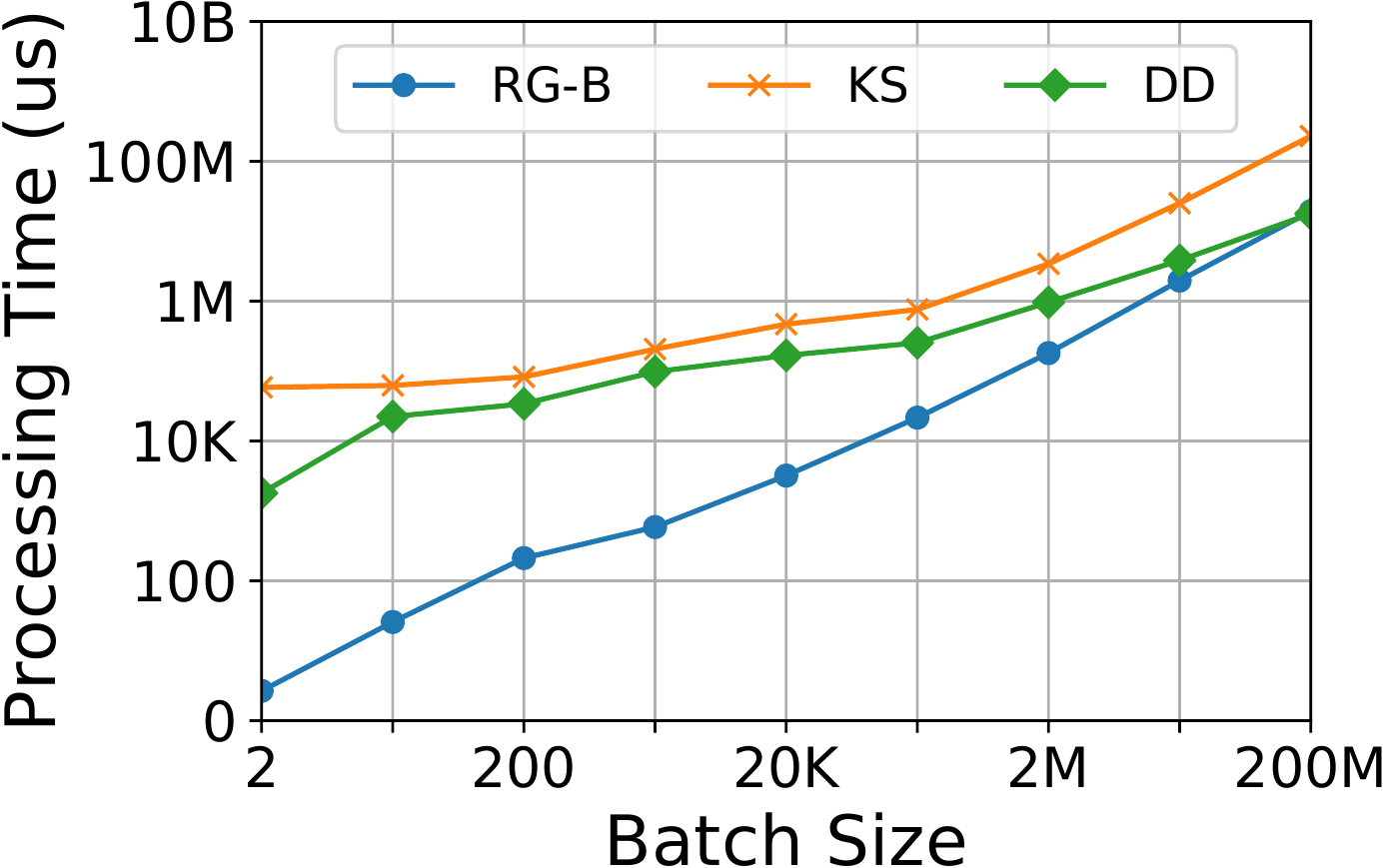}
    \subcaption{Latency of BFS on Twitter
    \label{fig:compare_kickstarter_latency}}
\end{minipage}
\begin{minipage}{0.245\linewidth}
	\centering
    \includegraphics[width=\columnwidth]{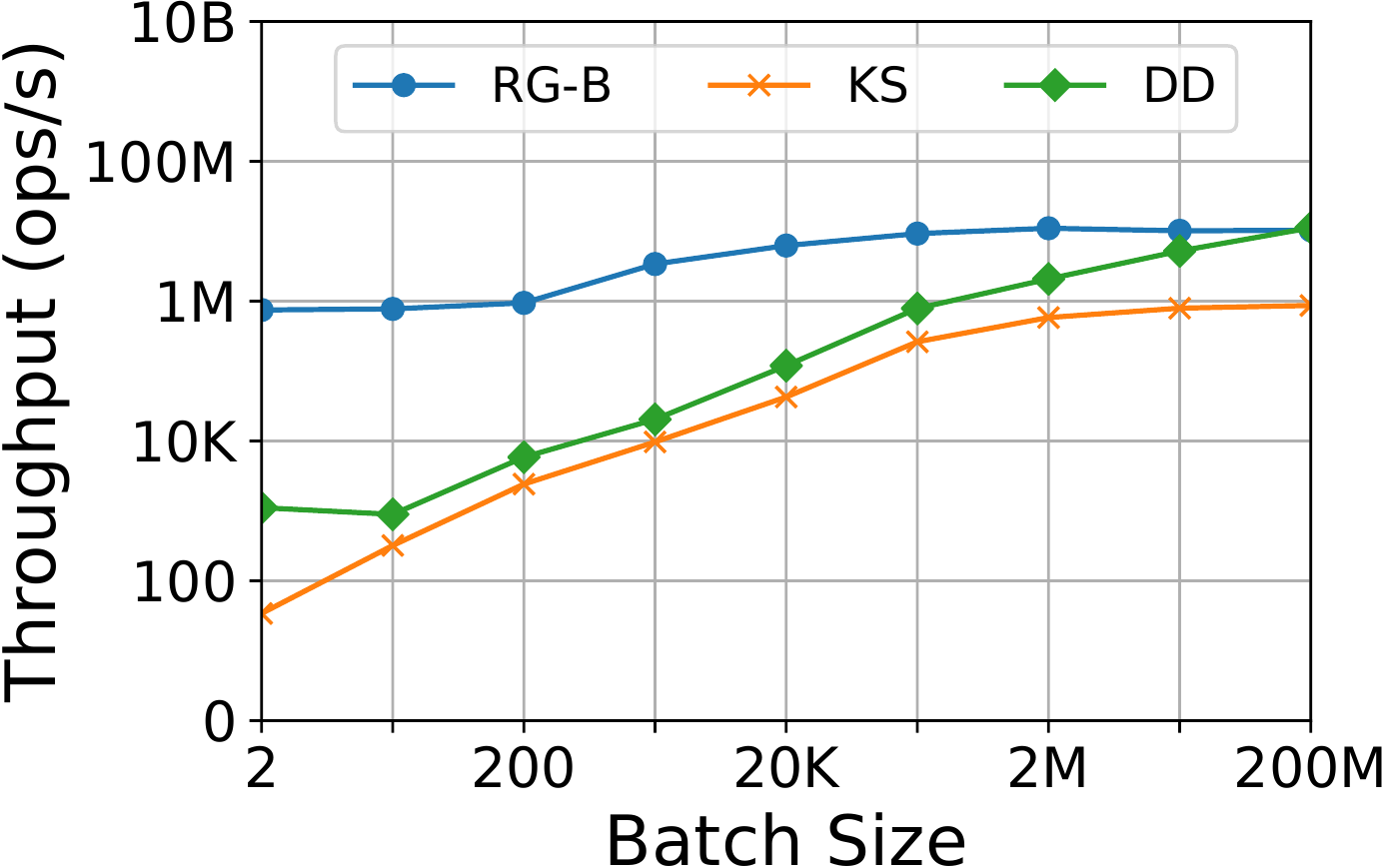}
    \subcaption{Throughput of BFS on Twitter
    \label{fig:compare_kickstarter_throughput}}
\end{minipage}
\vspace{-0.75em}
\caption{Performance of \SYS-Batch (RG-B), KickStarter (KS) and Differential Dataflow (DD) with different batch sizes}
\label{fig:compare_kickstarter}
\vspace{-1.5em}
\end{figure*}

\subsection{Comparisons with Existing Systems}
\label{subsec:comparisons}
We also evaluate the performance of \SYS compared with other streaming systems with batched updates. The goal is to evaluate the performance of \SYS when the scenarios allow batching updates together and also reducing analysis frequency. 
We choose KickStarter\footnote{A module of GraphBolt, is available at \url{https://bit.ly/34JgiX8}, commit 190d15a.} and Differential Dataflow\footnote{The latest implementation on Rust from \url{https://bit.ly/3hK8xnv}, commit 704bee.} as baselines in this evaluation.
Both of them officially provide BFS implementations, and we implement SSSP for Differential Dataflow based on its BFS code. Batch-update mode is enabled when it provides better performance.
In all implementations, vertex IDs are 64-bit integers to generally support large graphs which may contain more than 4 billion vertices. 
\SYS processes updates in batches and also disable WALs and tracing history for fair comparison.

We compare the performance of systems with different sizes of batches, from two updates (one edge insertion and one edge deletion) to 200M updates. The metric is the processing time of ingesting updates and performing analysis for each batch. And the throughput is calculated by the processing time and the batch size.

Figure~\ref{fig:compare_kickstarter_speedup} and Figure~\ref{fig:compare_differential_dataflow_speedup} show the geometric mean of speedups for all datasets and their distributions.
\SYS with batching outperforms KickStarter 13.8K times (from 587 with SSSP-HepPh to 588K with BFS-Bitcoin) and Differential Dataflow 1.06K times (from 365 with SSSP-HepPh to 4.45K with BFS-Bitcoin) on average when the batch-size is 2 (nearly per-update analysis). 

As the batch size increases, \SYS's advantage gradually decreases. \SYS keeps the advantage until the batch size is larger than 20M, but such a large batch seriously hurt the latency and analyzing frequency. Taking BFS on Twitter-2010 (Figure~\ref{fig:compare_kickstarter_latency} and Figure~\ref{fig:compare_kickstarter_throughput}) as an example, it takes GraphOne 0.76 s to re-compute BFS once, which is about \SYS's processing time on a batch of 2M updates. This example shows that incremental computing does not always optimize the analyzing performance when processing large batches, so it is reasonable to focus the design of \SYS on fine-grained updates and analyses.


Compared with KickStarter, \SYS's performance improvement mainly credits to our localized data access (Section~\ref{sec:opt_latency}).
\SYS outperforms Differential Dataflow primarily due to specialized graph-aware engine and incremental model.
For example, it takes Differential Dataflow 78$\times$ the processing time to re-compute BFS on Twitter-2010 compared to \SYS.

\section{Discussion}
\label{sec:discussion}

\paragraph{\bfseries Affected Areas Could Be Small.}
\label{subsec:affected_area}
\newcommand{\bbone}{\ensuremath{\vmathbb{1}}}

In our evaluation, \SYS can ingest up to millions of updates per second and provide per-update incremental analysis, thanks to localized data access and inter-update parallelism. Its performance surprises us and prompts us to study how each update modifies the results. We use \textit{affected area}~\cite{fan_incremental_2011,fan_incremental_2017} (AFF) as a tool to model computing costs. The affected area is the area modified or inspected (accessed) by an update. 

Incremental computing assumes that the affected area of each update is relatively small compared to the entire graph. Several incremental graph computing models~\cite{mcsherry_differential_2013,mahmood_tornado:_2015,sengupta_graphin:_2016,vora_kickstarter:_2017} can accelerate monotonic algorithms in practical, but they still lack sufficient discussion and analysis of affected areas.

We try to give some mathematical bounds of affected areas.
Because of the variety of graphs, updates, and algorithms, to mathematically bound affected areas for general monotonic algorithms is challenging. We only find a preliminary, yet optimistic bound based on an assumption. We assume that updating edges are uniformly sampled from all edges in the graph.

Consider a directed graph $G = (V, E)$ and a rooted tree in the graph $T = (r, V_T, E_T), r \in V_T \subset V, E_T \subset E$, in which edges point from parents to children. The rooted tree represents the dependency tree of monotonic algorithms. For any vertex $i \in V$, we define $T_i = (V_i, E_i)$, where $(V_i, E_i)$ is the subtree of $T$ rooted at $i$ if $i \in V_T$, and $(\varnothing, \varnothing)$ otherwise. The size of $T_i$ is denoted by $\vert T_i \vert = \vert V_i \vert$, satisfying $0 \le \vert T_i \vert \le \vert V_T \vert \le \vert V \vert$. 

For any directed edge $e = (i, j)$ in $G$, let $\text{AFFV}_e = \bbone_{e \in E_T} \vert T_j \vert$ and $\text{AFFE}_e = \bbone_{e \in E_T} \sum_{k \in V_{j}} d_k$ ($d_k$ is the total degree of $k$ in $G$, and $\bbone$ is the indicator function), respectively representing the upper bound 
of vertices that need to be modified after $e$ is inserted or removed, and edges related to these vertices. $\text{AFFV}_e$ bounds the area modified by $e$, and $\text{AFFE}_e$ bounds the area inspected by $e$.

When $e$ is sampled uniformly from the edge set $E$, the mean $\text{AFFV}$ can be written as:
\begin{align*}
\overline{\mathrm{AFFV}} &= \frac{1}{\vert E \vert} \sum_{e \in E} \mathrm{AFFV}_e  = \frac{1}{\vert E \vert} \sum_{e \in E} \bbone_{e \in E_T} \vert T_j \vert \\
& = \frac{1}{\vert E \vert} \sum_{e \in E_T} \sum_{k \in V_j} 1 = \frac{1}{\vert E \vert} \sum_{v \in V_T} (\depth_v + 1) \\
& \le \frac{\vert V_T \vert (D_T + 1)}{\vert E \vert} \le \frac{\vert V \vert}{\vert E \vert} (D_T + 1) = \frac{D_T + 1}{\overline{d}}
\end{align*}
where $\depth_v$ is the depth of $v$ in $T$ (the distance between $v$ and $r$), $D_T$ is the diameter (the length of the longest path) of $T$, and $\overline{d} \ge 1$ is the mean degree of $G$. So that $\overline{\mathrm{AFFV}}$ could be bounded by $D_T / \overline{d}$. 
In power-law graphs, $D_T$ is often small~\cite{milgram1967small,cohen_scale-free_2003,leskovec_graphs_2005,leskovec_kronecker_2010}. And in evolving graphs, $D_T$ will decrease while $\overline{d}$ will increase over time, according to \textit{Densification Laws} and \textit{Shrinking Diameters}~\cite{leskovec_graphs_2005}. 

Similarly, the mean $\text{AFFE}$ can be calculated as:
\begin{align*}
\overline{\mathrm{AFFE}} &= \frac{1}{\vert E \vert} \sum_{e \in E} \bbone_{e \in E_T}\sum_{k \in V_j} d_k = \frac{1}{\vert E \vert} \sum_{e \in E_T}\sum_{k \in V_j} d_k \\
&= \frac{1}{\vert E \vert} \sum_{v \in V_T} (\depth_v + 1) d_v \le \frac{1}{\vert E \vert} \sum_{v \in V_T} (D_T + 1) d_v \\
& \le \frac{D_T + 1}{\vert E \vert} \sum_{v \in V} d_v = \frac{D_T + 1}{\vert E \vert} \cdot 2 \vert E \vert = 2(D_T + 1)
\end{align*}
which can also be bounded by $D_T$. 

The results above guarantee mathematically that if we choose edges randomly for each update, only few vertices will be modified in average, and few vertices and edges will be accessed to perform the post-update analysis, thus showing the efficiency of incremental monotonic algorithms on power-law graphs.





\paragraph{\bfseries Performance with Non-power-law Graphs.}
\label{subsec:non-power-law}
Existing hierarchical algorithms~\cite{sanders_highway_2005,nannicini_shortest_2008,wang_querying_didi_2019} can efficiently limit the computing into a small neighbouring area for non-power-law graphs such as roadmaps. However, they cannot deal with power-law graphs efficiently because of hubs in power-law graphs. \SYS focus on taking one step forward to support power-law graphs efficiently based on incremental computing. 
Meanwhile, \SYS can also handle per-update incremental analysis on non-power-law graphs.

We evaluate \SYS with the USA road network~\cite{network_repository}, which is a non-power-law roadmap. There are 23.9M vertices and 28.9M edges in the USA dataset, and the experimental setup is the same as Section~\ref{subsec:setup}. The throughput is 26.7K ops/s for BFS, 4.10K ops/s for SSSP, 154K ops/s for SSWP, and 10.4K ops/s for WCC.

\section{Related Work}
\label{sec:related_work}
\paragraph{\bfseries Graph Computing on Static Graphs.}
A large number of systems~\cite{low_graphlab_2010,gonzalez_powergraph_2012,kyrola_graphchi_2012,shun_ligra_2013,gonzalez_graphx_2014,zhu_gridgraph_2015,zhu_gemini_2016,ai_squeezing_2017,maass_mosaic_2017,wang_graspan:_2017,shi_graph_gpu_2018,zhangyunming_graphit_2018,mukkara_exploiting_2018,shentu,chen_powerlyra_2019,vora_lumos_2019} focus on graph computing with static graphs. These systems are designed for efficient graph analytics on entire graphs, but they suffer from ETL overheads with evolving graphs.

\paragraph{\bfseries Dynamic Graph Stores.}
Graph databases~\cite{neo4j,titan,orientdb,arangodb,bronson2013tao,sun2015sqlgraph,triad,dubey2016weaver,carter2019nanosecond} mainly target transactional workloads, which rarely query and modify a large number of vertices or edges. Some recent work~\cite{jindal2015graph,kimura2017janus,case,all-in-one,zhu2019livegraph} propose to optimize analytical workloads in graph databases as well. 
Several graph computing engines~\cite{prabhakaran_managing_2012,han_chronos:_2014,miao_immortalgraph_2015,macko_llama_2015,iyer_time-evolving_2016,vora_synergistic_2016,then_automatic_2017,kumar_graphone:_2019} are also designed to support evolving graphs. 
These systems can handle graph update workloads. However, they are limited by recomputing on entire graphs for analytical workloads due to lack of incremental computing.

\paragraph{\bfseries Generalized Streaming Systems.}
Several generalized streaming systems, such as Storm~\cite{noauthor_storm_nodate} and Flink~\cite{katsifodimos_apacheflink_2016}, allow users to incrementally process unbounded streams. Spark Streaming~\cite{zaharia_discretized_2013} utilizes small batches to handle streaming data and provide second-scale latency. It is not easy for users to develop incremental iterative graph algorithms based on these systems, such as monotonic algorithms.
Differential Dataflow~\cite{mcsherry_differential_2013}, together with Naiad~\cite{murray_naiad:_2013}, carries out a generalized computational model that supports executing iterative and incremental computations with low latency. However, general-purpose systems lose opportunities to specialize and optimize for graph workloads.

\paragraph{\bfseries Graph Streaming Systems.}
Kineograph~\cite{cheng_kineograph:_2012} and GraphInc~\cite{cai_facilitating_2012} are systems that enable incremental computation for graph mining.  Qiu et al.~\cite{qiu_real-time_2018} design a real-time streaming system called GraphS for cycle detection. These systems lack support for monotonic algorithms like shortest path.
Tornado~\cite{mahmood_tornado:_2015} processes user queries by branching the execution and computing results incrementally while ingesting graph structure updates. However, it might lead to incorrect results in the presence of edge deletions when maintaining some monotonic algorithms such as WCC and SSWP~\cite{vora_kickstarter:_2017}.

KickStarter~\cite{vora_kickstarter:_2017} provides correct incremental computation for monotonic algorithms by tracing dependencies and trimmed approximations.
GraPU~\cite{sheng_grapu:_2018} accelerates batch-update monotonic algorithms by components-based classification and in-buffer precomputation.
GraphIn~\cite{sengupta_graphin:_2016} incorporates an I-GAS model that processes fixed-size batches of updates incrementally.
GraphBolt~\cite{mariappan_graphbolt:_2019} proposes a generalized incremental model to handle non-monotonic algorithms like Belief Propagation, but involves more overheads than KickStarter for monotonic algorithms.
These systems support monotonic algorithms, but all of them are designed for batch-update analysis. \SYS targets per-update analysis to provide milliseconds tail-latency and detailed information in comparison.


\section{Conclusion}

In this paper, we present \SYS, a real-time streaming system that efficiently supports per-update incremental analysis for monotonic algorithms on evolving graphs.
The main ideas of \SYS are localized data access and inter-update parallelism, which are critical to achieve high throughput and low latency simultaneously.

\begin{acks}
We sincerely thank all SIGMOD reviewers for their insightful comments and suggestions. 
We also appreciate suggestions from Yaqin Li, Yanzheng Cai, Xuanhe Zhou, Bowen Yu, Songtao Yang, Jidong Zhai, Xiaosong Ma, and Marco Serafini.
This work is partially supported by NSFC 61525202 and BAAI scholar program. Corresponding authors are Wentao Han and Wenguang Chen.
\end{acks}

\bibliographystyle{ACM-Reference-Format}
\bibliography{reference}

\end{document}